\documentclass[fleqn,usenatbib, hyperref={colorlinks = true,linkcolor = blue, allcolors=blue}]{mnras}

\usepackage{newtxtext,newtxmath}
\usepackage[T1]{fontenc}
\DeclareRobustCommand{\VAN}[3]{#2}
\let\VANthebibliography\thebibliography
\def\thebibliography{\DeclareRobustCommand{\VAN}[3]{##3}\VANthebibliography}

\usepackage{graphicx}	
\usepackage{amsmath}	
\usepackage{gensymb}

\usepackage{booktabs}
\usepackage{varwidth}

\newdimen\digitwidth    
\setbox0=\hbox{\rm0}
\digitwidth=\wd0
\catcode`!=\active
\def!{\kern\digitwidth}
\normalsize

\usepackage{color}
\usepackage{colordvi}
\usepackage{ulem}
\RequirePackage{lineno}

\title[Five FRBs discovered in the FAST GPPS survey]
{The FAST Galactic Plane Pulsar Snapshot survey:
  \\ IV. Discovery of five fast radio bursts}

\author[Zhou et al.]
{D.~J. Zhou$^{1,2}$,
        J.~L. Han$^{1,2,3}$\thanks{E-mail: hjl@nao.cas.cn (JLH)},
        W.~C. Jing $^{1,2}$,
        P.~F. Wang$^{1,2}$,
        C. Wang$^{1,2}$,
        T. Wang$^{1,2}$,
        W.-Y. Wang$^{2,4,5}$,
        \and
        R. Luo$^{6,7}$,
        J. Xu$^{1}$,
        R.~X. Xu$^{8}$,
        H.~G. Wang$^{7,9}$
         \\
         $^1$National Astronomical Observatories, Chinese Academy of Sciences,
         Jia-20 Datun Road, ChaoYang District, Beijing 100101, China\\
         $^2$School of Astronomy, University of Chinese Academy of Sciences,
         Beijing 100049, China  \\
$^{3}$CAS Key Laboratory of FAST, NAOC, Chinese Academy of Sciences, Beijing 100101, China\\
$^{4}$School of Physics and State Key Laboratory of Nuclear Physics and Technology, Peking University, Beijing 100871, China\\
$^{5}$Kavli Institute for Astronomy and Astrophysics, Peking University, Beijing 100871, China\\
$^{6}$CSIRO Space and Astronomy, PO Box 76, Epping, NSW 1710, Australia\\
$^{7}$Department of Astronomy, School of Physics and Materials Science, Guangzhou University, Guangzhou 510006, China\\
$^{8}$Department of Astronomy, Peking University, Beijing 100871, China\\
$^{9}$National Astronomical Data Center, Great Bay Area, Guangzhou 510006, China
}

\date{Accepted XXX. Received YYY; in original form ZZZ}

\pubyear{2022}

\begin{document}
\label{firstpage}
\pagerange{\pageref{firstpage}--\pageref{lastpage}}
\maketitle

\begin{abstract}
We report five new fast radio bursts (FRBs)  discovered from the Galactic Plane Pulsar Snapshot (GPPS) survey by the Five-hundred-meter Aperture Spherical radio Telescope (FAST): FRB\,20210126, FRB\,20210208, FRB\,20210705,  FRB\,20211005 and FRB\,20220306. To date, no repeating bursts from these FRB sources have been detected in the follow-up monitoring observations, leading to their classification as potential one-off events. We obtain the basic parameters for these bursts, including position, dispersion measure (DM), pulse width, spectral index, scattering time-scale, etc. The fluences and flux densities are generally lower in comparison to the values observed in one-off bursts discovered by other telescopes. Among the observed bursts, polarization data for 4 bursts were recorded during observations. Consequently, we obtain polarization profiles and Faraday rotation measures (RMs) for these bursts.  
\end{abstract}



\section{Introduction}

Since the discovery of the first fast radio burst (FRB) by \citet{Lorimer2007Sci}, hundreds of FRBs have been reported\footnote{\url{https://www.herta-experiment.org/frbstats/catalogue}\label{fn:frbcat}}, with the majority being one-off events \citep{CHIME2021ApJS}, and a few dozen are repeaters ~\citep{Spitler2016Natur, CHIME2019ApJ...885L..24C, Fonseca2020ApJ...891L...6F, 2023arXiv230108762T}. Notable among the repeaters are FRB 20121102A, which was the first repeater identified \citep{Spitler2016Natur}, and FRB 20201124A, which displays heightened activity during specific episodes \citep{Xu2022Natur, ZhouDJ2022RAA}. Interferometric observations enable good localization of repeaters, allowing for precise identification of its host galaxy. It is important to note that all FRBs originate in extragalactic sources, as evidenced by their dispersion measures (DMs) greatly exceeding the DM upper limits associated with the Milky Way. One-off sources, characterized by a single detected pulse, pose challenges in terms of positional determination and follow-up investigations of their source properties.

In contrast to pulsars which are predominantly found in the Galactic plane,  FRBs are typically discovered in an isotropic manner across the celestial sphere, depending upon the observable sky regions of the telescopes utilized. An in-depth analysis of the CHIME dataset reveals no discernible evidence supporting a correlation between the distribution of FRBs and the latitude of the Galaxy \citep{CHIME2021ApJS,Josephy2021ApJ}. Notably, FRBs identified at low Galactic latitudes ($|b|\le10\degree$) have been identified through a single-pulse search conducted on data from pulsar surveys \citep{BurkeSpolaor2014ApJ, Petroff2014ApJ}. As of now, more than 80 one-off FRB sources have been detected in the low Galactic latitude region. 

Though the emission mechanism of FRBs is not known yet, their cosmological origin is indicated by their large DMs. To unravel their enigma, sensitive observations are obviously necessary to discover more bursts and measure their properties.  Currently, the Five-hundred-meter Aperture Spherical radio Telescope \citep[FAST,][]{Nan2006ScChG,Nan2011IJMPD}, equipped with the 19-beam L-band receiver \citep{JiangP2020RAA}, is currently the most sensitive single-dish radio telescope at the L-band. It has been judiciously used to reveal the detailed properties of several repeaters. Distinct variations in polarization angle curves for bursts have been detected in the case of FRB 20180301A \citep{Luo2020Natur}, while strikingly unexpected fluctuations in Faraday rotation measures (RMs) have been unveiled from FRB 20201124A \citep{Xu2022Natur}.  A large number of bursts have been detected from both FRB 20121102A \citep{LiD2021Natur} and FRB 20201124A \citep{Xu2022Natur,ZhouDJ2022RAA}. The latter source has undergone exhaustive analyses of burst morphologies \citep{ZhouDJ2022RAA}, energy distribution \citep{ZhangYK2022RAA}, polarization properties \citep{JiangJC2022RAA} and even spin period exploration \citep{NiuJR2022RAA}. 

Among the numerous sources where only a single burst has been detected, approximately a dozen have had their polarization data recorded during the discovery sessions. These data unveil the polarization properties and RMs of these sources. Typically, these FRBs exhibit RMs of several hundred rad\,m$^{-2}$. The degree of circular polarization is generally less than 30\%, and the degree of linear polarization varies from 0\% to nearly 100\%.

\begin{table}
\centering
\caption{FAST observation sessions for five newly discovered FRBs}
\label{tab1}
\setlength{\tabcolsep}{9.0pt}
\footnotesize
\begin{tabular}{lccrr}
 \hline\noalign{\smallskip}
 Name        & Date     & MJD   & T$_{\rm obs}$    & Burst \\
              &          &       & (min)            & No.   \\
 \hline  
 FRB\,20210126 & 20210126 & 59240 & 15               & 1     \\
 (J1905+0941)  & 20220607 & 59737 & 30               & 0     \\
              & 20220529 & 59728 & 15               & 0     \\
              & 20220824 & 59815 & 40               & 0     \\
              & 20221210 & 59923 & 20               & 0     \\
              & 20230105 & 59949 & 40               & 0     \\
              & 20230117 & 59961 & 40               & 0     \\
              & 20230408 & 60042 & 40               & 0     \\
              & 20230422 & 60056 & 40               & 0     \\[2mm]
 FRB\,20210208 & 20210208 & 59253 & 5                & 1     \\%
(J1902+1324)   & 20210826 & 59452 & 15               & 0     \\
              & 20210903 & 59460 & 30               & 0     \\
              & 20211018 & 59505 & 60               & 0     \\
              & 20220824 & 59815 & 40               & 0     \\
              & 20230117 & 59961 & 40               & 0     \\
              & 20230408 & 60042 & 40               & 0     \\
              & 20230422 & 60056 & 40               & 0     \\[2mm]
 FRB\,20210705 & 20210705 & 59400 & 5                & 1     \\%
 (J2012+3858)  & 20210826 & 59452 & 15               & 0     \\
              & 20210903 & 59460 & 30               & 0     \\
              & 20211018 & 59505 & 60               & 0     \\
              & 20221022 & 59874 & 40               & 0     \\
              & 20230105 & 59949 & 40               & 0     \\
              & 20230117 & 59961 & 40               & 0     \\
              & 20230408 & 60042 & 40               & 0     \\
              & 20230422 & 60056 & 40               & 0     \\[2mm]
 FRB\,20211005 & 20211005 & 59492 & 15               & 1     \\%
(J1945+2407)  & 20220406 & 59675 & 60               & 0     \\
              & 20220409 & 59678 & 60               & 0     \\
              & 20220627 & 59757 & 60               & 0     \\
              & 20221104 & 59887 & 40               & 0     \\
              & 20230105 & 59949 & 40               & 0     \\
              & 20230117 & 59961 & 40               & 0     \\
              & 20230408 & 60042 & 40               & 0     \\
              & 20230422 & 60056 & 40               & 0     \\[2mm]
 FRB\,20220306& 20220306 & 59644 & 15               & 1     \\%
 (J1946+2259) & 20221109 & 59892 & 15               & 0     \\
              & 20230204 & 59979 & 60               & 0     \\
              & 20230210 & 59985 & 60               & 0     \\
  \hline
  \end{tabular} 
  \end{table}

\begin{figure*}
    \centering
    \includegraphics[width=0.33\textwidth]{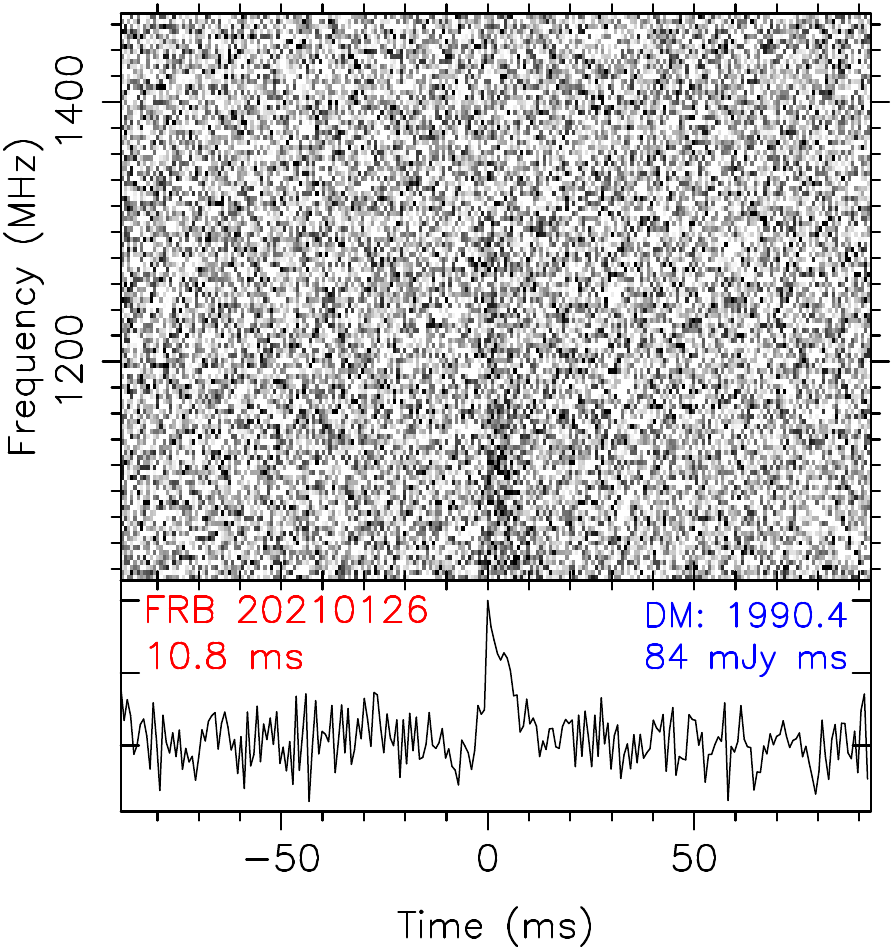} 
    \includegraphics[width=0.33\textwidth]{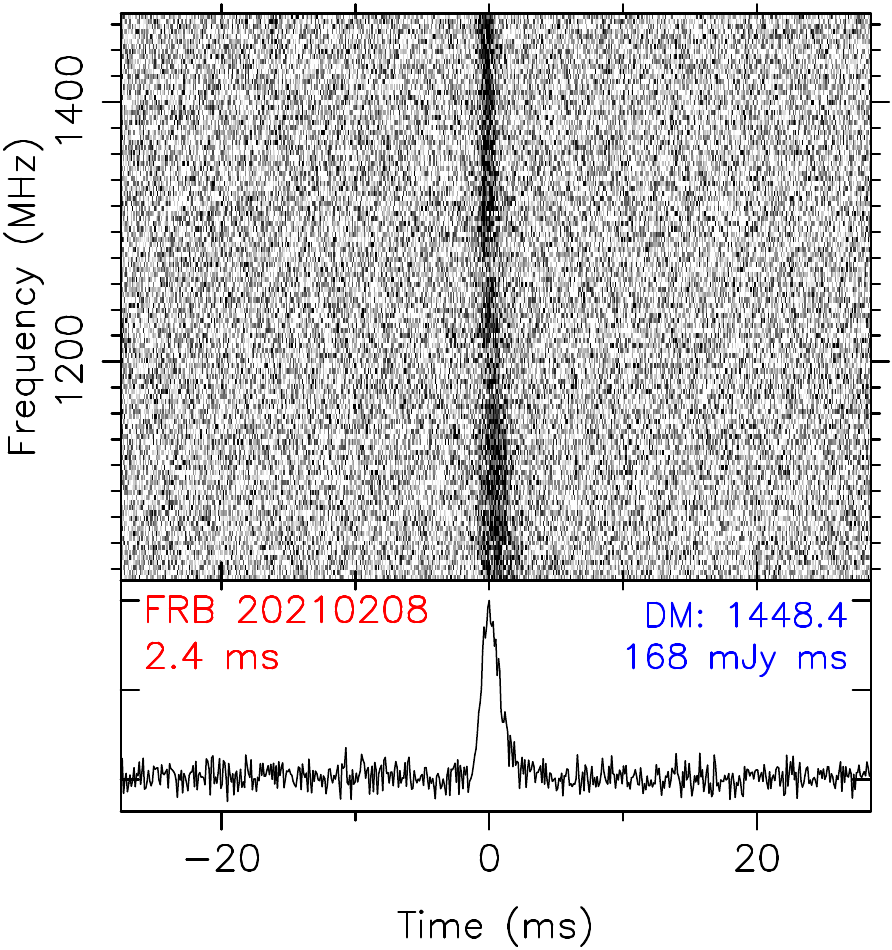} 
    \includegraphics[width=0.33\textwidth]{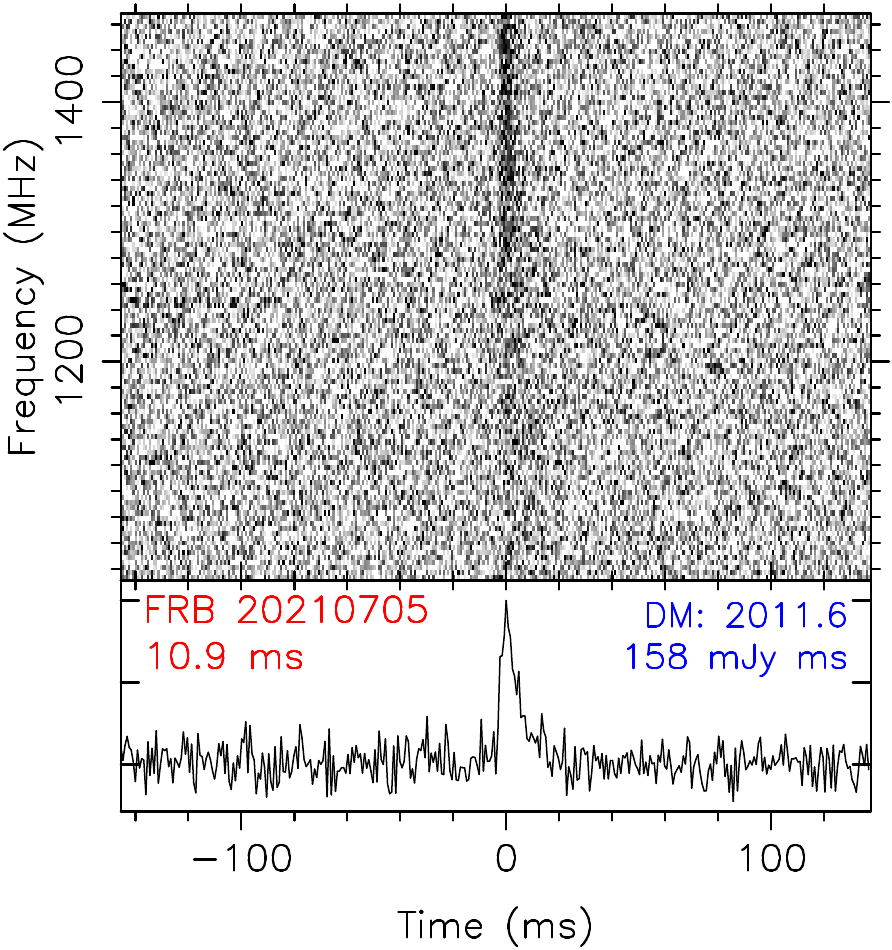} 
    \includegraphics[width=0.33\textwidth]{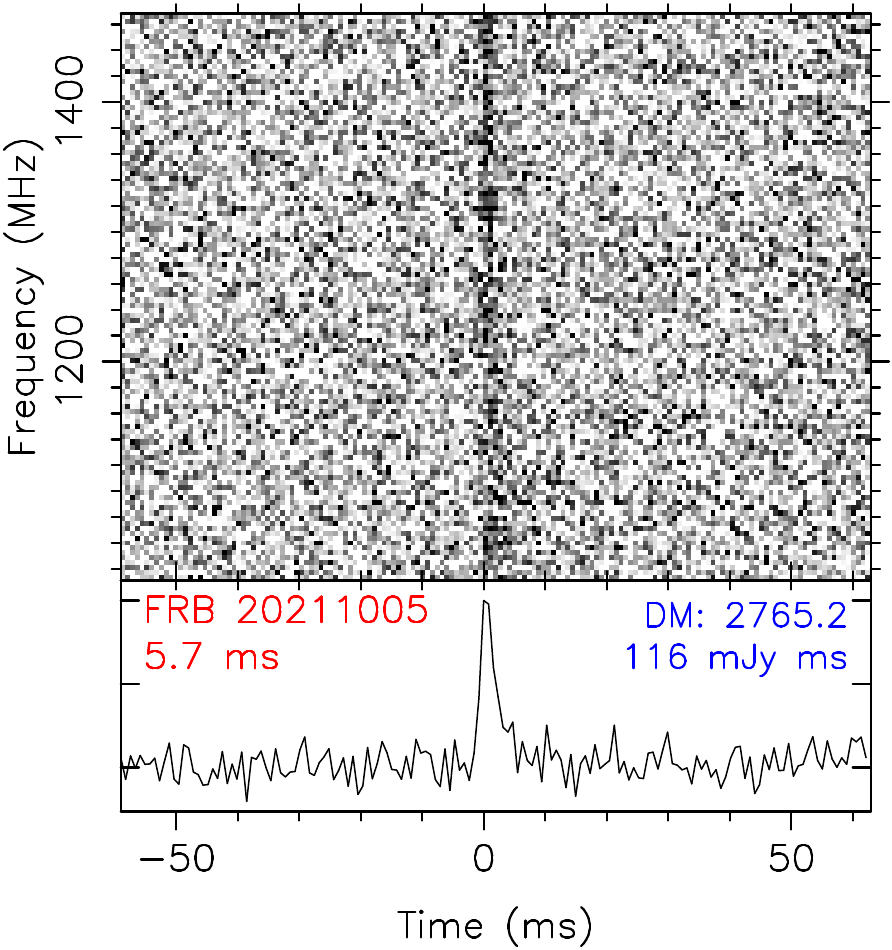} 
    \includegraphics[width=0.33\textwidth]{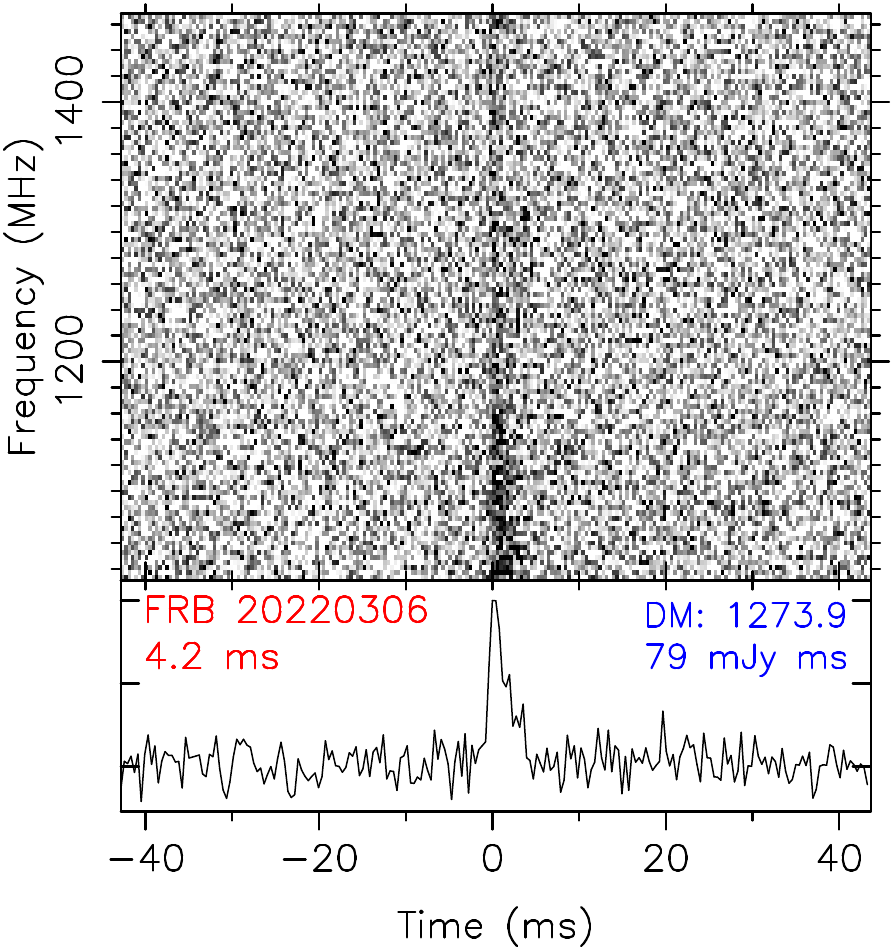} 
    \caption{The dynamic spectra for five FRBs discovered in the FAST GPPS survey are shown in the upper sub-panel, and the averaged pulse profile in the bottom sub-panel. The FRB names and pulse width in ms are marked on the left, and the DM and the total fluence on the right.}
    \label{fig:TFplot}
\end{figure*}

\begin{table*}
\centering
\caption{Basic parameters of the newly discovered FRBs in the FAST GPPS survey}
\label{tab:sourcelist}
\setlength{\tabcolsep}{7.0pt}
\footnotesize
\begin{tabular}{lccccc}
\hline
Name                & FRB\,20210126       & FRB\,20210208       & FRB\,20210705       & FRB\,20211005       & FRB\,20220306       \\
                    & (J1905+0941)        & (J1902+1324)        & (J2012+3858)  & 
(J1945+2407)   &  (J1946+2259) 
\\
\hline
T$_{\rm obs}$ (min) & 280                & 270                & 310                & 355                & 150                    \\
R.A. (hh:mm:ss)*     & 19:05:37.1        & 19:02:49.3        & 20:12:26.8        & 19:45:22.4        & 19:46:45.0        \\
Dec ($\pm$dd:mm:ss)* & +09:41:05         & +13:24:53         & +38:58:29         & +24:07:56         & +22:59:14         \\
Galactic: $l$  (\degree)       & 43.1595           & 46.1672           & 76.0306           & 60.3770           & 59.5450           \\
Galactic: $b$  (\degree)       & 1.2870            & 3.6008            & 2.7112            & $-$0.1694           & $-$1.0170           \\
T$_{\rm sys}$ (K)  & 29.0  & 26.7  & 28.1  & 29.5  &  28.5\\
DM  (pc~cm$^{-3}$)   & 1990.4$\pm$2.8    & 1448.4$\pm$0.7    & 2011.6$\pm$3.2    & 2765.2$\pm$1.9    & 1273.9$\pm$0.9    \\
DM$\rm_{MW}$ (pc~cm$^{-3}$)** & 724 / 717   & 432 / 425    & 484 / 334        & 528 / 512         & 454 / 505         \\
DM$\rm _{E}$ (pc~cm$^{-3}$)** & 1266 / 1273 & 1016 / 1023  & 1528 / 1678      & 2237 / 2253       & 820 / 769         \\
TOA (MJD) ***           & 59240.145608646   & 59253.114591878   & 59400.809573843   & 59492.509062134   & 59644.118164808   \\
W$_{\rm obs}$ (ms)  & 10.8$\pm$0.8               & 2.4$\pm$0.1               & 10.9$\pm$0.8              & 5.7$\pm$0.8               & 4.2$\pm$0.4               \\
Intrinsic $ W_{\rm 1GHz}$ (ms) &   $9.2_{-5.2}^{+6.7}$   &$3.6_{-0.3}^{+0.3}$&      $22.7_{-8.2}^{+7.8}$        &$5.1_{-2.6}^{+3.8}$&$3.5_{-0.8}^{+0.9}$\\
Scattering $\rm{\tau}_{1 GHz}$ (ms) & $8\pm2$ & $1.3\pm0.1$ & $12\pm2$ & $3\pm1$ & $2.9\pm0.5$ \\
Width-evolving index $\beta$  &   $-2.5_{-1.8}^{+3.2}$   &$-2.0_{-0.3}^{+0.3}$&$-4.0_{-0.7}^{+1.3}$&$-1.5_{-1.6}^{+1.6}$&$-3.8_{-0.8}^{+1.4}$\\
Peak flux density $S_{\rm peak}$ (mJy)& 19.3$\pm$2.8          & 103.8$\pm$5.7   & 23.3$\pm$2.1   & 29.0$\pm$2.2    & 33.3$\pm$3.0              \\
Fluence $F_\nu$ (mJy ms)    & 83.8$\pm$6.2    & 167.8$\pm$3.2         & 157.7$\pm$6.6     & 115.6$\pm$5.8          & 78.7$\pm$3.8                \\
Spectral index $\alpha$ & $-$6.1$\pm$3.5  & $-$2.2$\pm$0.1  & 2.0$\pm$0.6      & $-$0.7$\pm$0.7      & $-$6.9$\pm$1.1       \\
$L/I$ (\%)          &  --                 & 12.0$\pm$2.9              & 61.1$\pm$9.0              & 76.6$\pm$7.2              & 74.8$\pm$6.4              \\
$V/I$ (\%)          &  --                 & 6.9$\pm$3.4               & $-$6.6$\pm$5.5        &     $-$11.7$\pm$5.0             & $-$17.1$\pm$5.9             \\
$|V|/I$ (\%)        &  --                 & 5.8$\pm$1.5               & 5.2$\pm$1.5               & 10.1$\pm$2.1              & 16.8$\pm$3.1              \\
RM (rad m$^{-2}$)   &  --                 & 405$\pm$6            & $-$407$\pm$8           & $-$1739$\pm$4          & 86$\pm$6             \\
$\rm RM_{E}$ (rad m$^{-2}$) & --          & $-$144$\pm$72          & $-$260$\pm$61          & $-$1661$\pm$30         & 146$\pm$30           \\
\hline
\multicolumn{6}{l}{Note: * The position has an uncertainty of $1.5'$. ** Two values were estimated respectively by using the two models for the Galactic electron}\\
\multicolumn{6}{l}{ density distribution: NE2001 / YMW16. 
*** The barycentric peak time at infinite frequency.}
\end{tabular}
\end{table*}

Undoubtedly, the detection capability of faint FRBs at greater distances is notably enhanced by the employment of a sensitive single-dish telescope endowed with a larger aperture \citep{Zhang2018ApJ}. Recently, the Commensal Radio Astronomy FAST Survey (CRAFTS) has discovered four faint and distant one-off FRBs \citep{ZhuWW2020ApJ, NiuxCH2021ApJ} and an active repeater \citep{NiuCH2022Natur}. These results highlight the capability of FAST to find new FRBs.

In this paper, we present the discovery of 5 FRBs, as detailed in Table~\ref{tab1}. These FRBs were detected as part of the FAST Galactic Plane Pulsar Snapshot (GPPS) survey\footnote{\url{http://zmtt.bao.ac.cn/GPPS/}} \citep{Han2021RAA}. The primary objective of the GPPS survey is to hunt for pulsars within the Galactic latitude range of $|b|\le10\degree$ in the observable sky region of the FAST telescope.  The survey has already discovered over 600 faint pulsars \citep{Han2021RAA, zhou2023RAA}.  In Section~\ref{2-obs}, we  briefly introduce the observations and data processing. In Section~\ref{3-results}, we present results for the 5 newly discovered FRBs. Finally,  discussion is presented in section~\ref{4-conc}.

\section{The GPPS survey, follow-up Observations and data processing}
\label{2-obs}

The FAST GPPS survey \citep[see details in][]{Han2021RAA} observes the sky region near the Galactic plane by using the L-band 19-beam receiver that covers the frequency range of 1000 -- 1500 MHz \citep{JiangP2020RAA}. The system temperature of the receiver is around 22K. For each of the 19 beams with a beam size of approximately $3'$, data from 2048 frequency channels within this range are recorded, with a sampling time of 49.152~$\mu$s. All observations are carried out within the zenith angles smaller than 28.5$^{\circ}$, so that the full gain of the FAST telescope is obtained from a full illumination area of 300~m in diameter from the primary reflector \citep{Nan2006ScChG}. The survey employs the {\it snapshot} mode within which the 19 beams with a separation of about $6'$ are well-organized in four pointings, so that successively shifts of 19 beams can fully cover a hexagonal sky region of 0.1575 square degrees. Each pointing lasts for 5 minutes for integration.

Subsequent to the initial detection of new pulsars or single pulses from any novel sources, follow-up tracking observations are commonly conducted to validate the findings. These tracking observations typically span a duration of 15 minutes. In specific instances within FAST projects, such as those focused on investigating rotating radio transients or FRBs, significantly longer observation times are taken. During these observation sessions, the central beam of the L-band 19-beam receiver is aligned with the initial detection coordinates. Comprehensive polarization data, including XX, X$^*$Y, XY$^*$ and YY, are recorded from all 19 beams.

In the data processing stage of the FAST GPPS survey, the XX and YY data undergo scaling based on the root mean square (RMS) of each channel. At the two edges of the observation band, the receiver gain decreases significantly, and hence data of 128 channels (31.25 MHz) each side are discarded. These scaled data sets are then combined to create a new FITS file in the "data1j" repository dedicated to pulsar searching, as detailed by Han et al. (2021) \citep{Han2021RAA}. For transient sources, we have developed novel source code that facilitates rapid incoherent dedispersion. This code generates DM-time images utilizing graphics processing units (GPUs), spanning the DM range of 3 to 3700 pc cm$^{-3}$, with a DM step size of 1 pc cm$^{-3}$. To optimize computational efficiency and enhance detection sensitivity for single pulses with a pulse width typically exceeding 1 ms, the data is down-sampled from 49.152 $\mu$s to $4 \times 49.152 {\rm \mu s} = 196.608$ $\mu$s. However, it is important to note that the combination of downsampling and the dispersion step size renders the single-pulse search method less attuned to narrow pulses lasting less than 1 ms. See details in the GPPS paper II \citep{zhou2023RAA}.

Utilizing data from the FAST GPPS survey and other FAST PI projects obtained by the 19-beam L-band receivers, data of a total of 17747 beam*hours are accessible for the purpose of FRB finding, as presented here. The FRBs can be recognized as  distinctive features in the DM-time images using the \texttt{YOLO} object detection technology developed in the  \texttt{Darknet}\footnote{\url{https://pjreddie.com/darknet/}} neural network framework. See \citet{zhou2023RAA} for details. Within such a large dataset, we successfully detected 5 FRBs. Among these, 2 FRBs were detected during the snapshot survey observations with a 5-minute pointing duration. The remaining 3 FRBs were identified from the follow-up verification tracking observations with a duration of 15 minutes. Table~\ref{tab1} provides details on the discovery observations for these FRBs and also specific follow-up observations.

\section{The properties of 5 FRBs}
\label{3-results}

Figure~\ref{fig:TFplot} presents the dynamic spectra and integrated frequency profiles pertaining to the 5 identified FRBs. Following the initial detection of a burst, a sequence of follow-up observations were conducted, the specifics of which are comprehensively outlined in Table~\ref{tab1}. No additional bursts were detected during these follow-up observation sessions, suggesting that these FRBs are likely either one-off events, or alternatively, from repeaters characterized by very low repetition rates.

The utilization of a multiple-beam receiver facilitates the determination of positions for FRBs. The beam size is about 3 arcminutes in diameter for the FAST L-band 19-beam receiver, and the separation between adjacent beams is approximately 6 arcminutes. 
Given that a burst is detected within a singular beam, the beam center is taken as the position of an FRB with a formal uncertainty of $\sigma_{\rm pos}\sim$ 1.5 arcminutes, i.e. the half of the beam size. Such a one-off burst was detected in one, not in the other beam 6 arcminutes away. If it were located at 2$\sigma_{\rm pos}$ away, it would be detected by other beams of the L-band 19-beam receiver.  

A comprehensive analysis of the 5 identified FRBs has been conducted, yielding fundamental parameters in Table~\ref{tab:sourcelist}. Detailed analysis and discussions are presented in the following.

\subsection{DMs}

To extract the optimal DM for each burst, we follow the approach outlined by \citet{ZhuWW2020ApJ}. Initially, in the vicinity of the initial detection time and DM associated with an FRB, we construct a two-dimensional representation depicting the variation of integrated pulse energy concerning both time and DM. This representation is obtained through the squared Gaussian-smoothed profile. Subsequently, we perform an integration of the pulse energy over a narrow time interval to derive the derivative concerning time. Finally, by fitting the resulting peak, we ascertain the DM value that yields the highest optimization for each burst. The degree of uncertainty is established at half of the value obtained from the fitting of the peak.

\begin{figure}
    \centering
    \includegraphics[width=0.95\columnwidth]{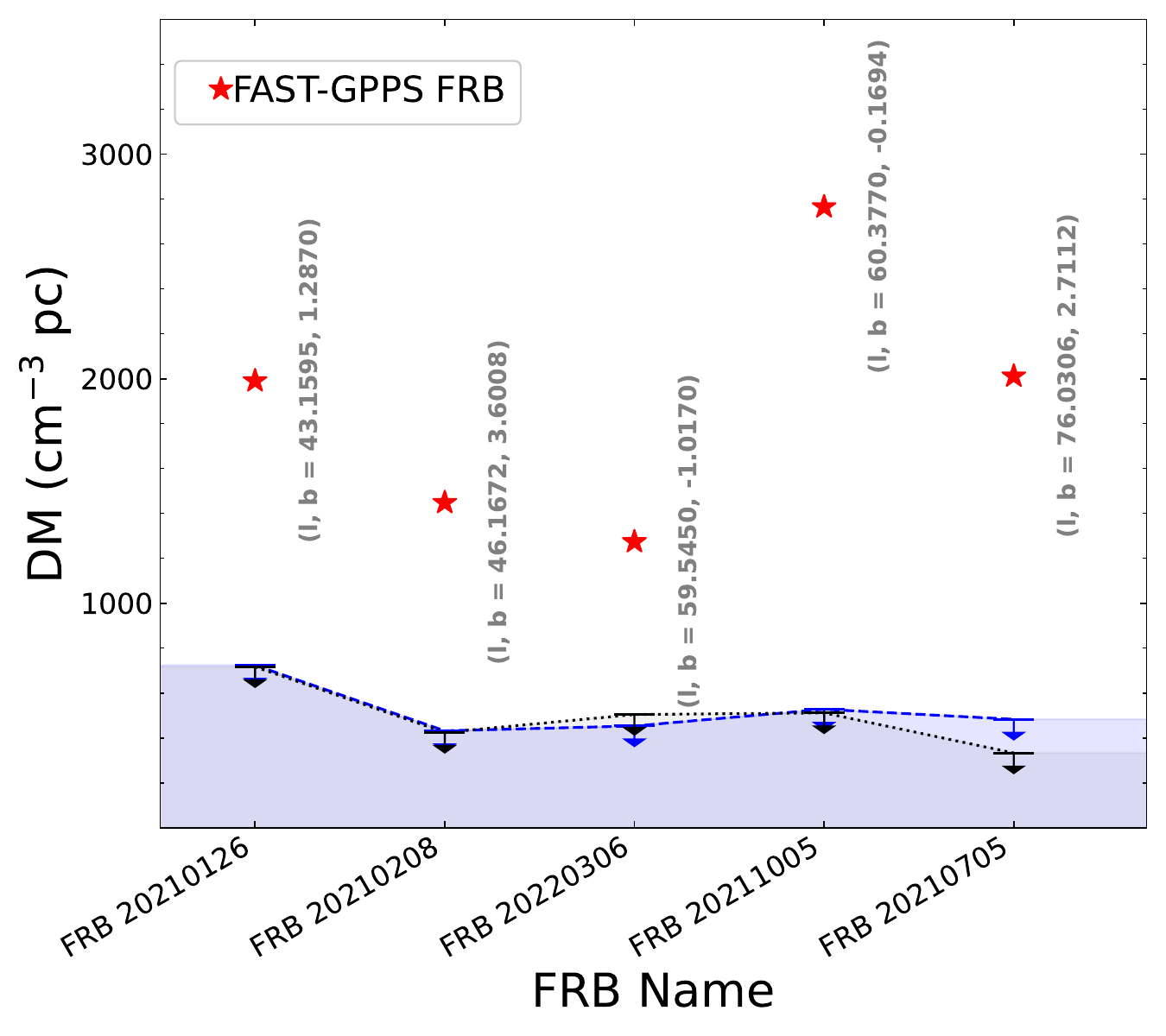}
    \caption{The DMs of these 5 FRBs remarkably exceed the DM upper limit of the Milky Way predicted by the electronic distribution models for the Milky Way, namely NE2001 (dashed line) and/or YMW16 (dotted line) \citep{NE2001,YMW2016}.}
    \label{fig:dm-name}
\end{figure}

The observed DMs for the five FRBs are outlined in Table~\ref{tab:sourcelist}. A comparison is made between these observed DMs and the maximum DMs predicted by the  electronic distribution in the Milky Way under two models, namely the NE2001 and YMW2016 models \citep{NE2001,YMW2016}. This comparison is illustrated in Figure~\ref{fig:dm-name}. Though the electronic distribution models might lack precision particularly in regions with low Galactic latitudes, causing large uncertainties in these lines of sight, the observed DMs of the five FRBs remarkably surpass the predicted maximum DMs associated with the Milky Way, implying the extragalactic origin for these bursts. The excess DM ($DM_{\rm E}$) can be reasonably attributed to extragalactic contribution, including the contributions from the host galaxy and intergalactic medium. 
 
Out of the observed FRBs, FRB\,20211005 exhibits the most considerable excess DM, reaching up to $\rm DM_{E} \sim $ 2237 $\rm pc\,cm^{-3}$ or 2253 $\rm pc\,cm^{-3}$. An approach to estimating the redshift ($z$) involves assuming that the excess DM originates from both the intergalactic medium and the host galaxy, the latter of which could potentially contribute as much as our own Milky Way. In this context, the estimation of $z$ can be accomplished using the formula provided by \citep{Deng2014ApJ}:
\begin{equation}
{DM_{\rm IGM}} = \frac{3cH_0\Omega_{\rm b}f_{\rm IGM}}{8\pi Gm_{\rm p}}\int^z_0\frac{\chi(z)(1+z)dz}{[\Omega_{\rm m}(1+z)^3+\Omega_{\Lambda}]^{\frac{1}{2}}},
\label{eq:DMIGM}
\end{equation} 
where the free electron number per baryon in the universe, denoted as $\chi(z)$, is approximately 7/8. Additionally, the fraction of baryons in the intergalactic medium is around $f_{\rm IGM}\sim0.83$, as indicated by other parameters provided in the work of \citet{Planck2020AA}. If we exclude the contribution from the host galaxy ($DM_{\rm host}$) which is assumed to be the same as that from our Galaxy, we can estimate that the intergalactic medium contributes a dispersion measure of $DM_{\rm IGM}=$1709$\rm\,pc\,cm^{-3}$ or 1741$\rm\,pc\,cm^{-3}$. Consequently, we derive that the redshift $z$ of the FRB\,20211005 event is approximately 1.94 or 2.00.

\begin{figure}
    \centering
    \includegraphics[width=0.95\columnwidth]{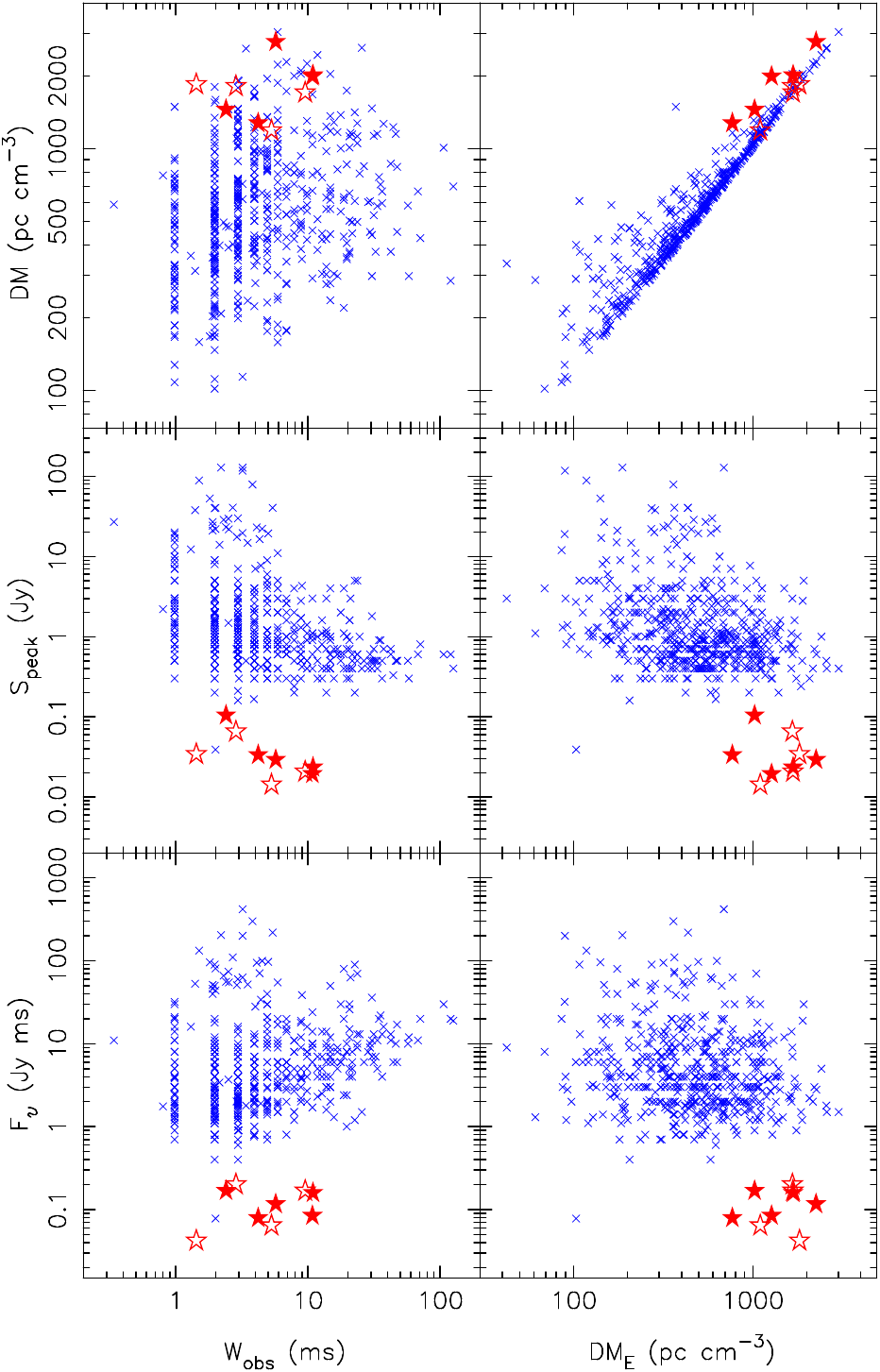}
    \caption{Basic parameters for FRBs discovered by FAST in the GPPS survey (solid star) and the CRAFTS ~\citep[hollow star,][]{ZhuWW2020ApJ,NiuxCH2021ApJ} and for other known one-off events (cross) detected above 400 MHz\footref{fn:frbcat}, mostly by \citet{CHIME2021ApJS}. The $DM_{\rm E}$ is estimated by using the YMW16 model~\citep{YMW2016}.}
    \label{fig:wf}
\end{figure}

\subsection{Time of arrival}

Using the optimal DM value determined early, we dedisperse and the pulse, resulting in the integrated profile over the observing frequency band. The time of arrival (TOA) for each burst is defined as the time of emission peak at an infinite frequency by  considering the best-fitted DM. Subsequently, these TOA values are converted to the Solar barycentric peak time using the DE438 ephemeris, as listed in  Table~\ref{tab:sourcelist}. Note here that the small uncertainties in both the DM and the source position might introduce a minor deviation of the true TOA value.

\subsection{Pulse-width and scattering} \label{subsec:scat}

To determine the pulse width, a method involving the application of box-car filters with differing widths to the pulse profile is employed. This procedure generates a graph that plots the maximum value of the matched-filtered profile against the width of the box-car filter. The observed pulse width ($W_{\rm obs}$) is represented by the peak of this graph, with an uncertainty equivalent to one data bin size which is 196.608 $\mu$s here. Furthermore, the signal-to-noise ratio (S/N) of the pulse is computed by dividing the peak value of this graph by the standard deviation ($\sigma$) of the adjacent off-pulse region, as outlined in the work by \citet{zhou2023RAA}.

The lower frequency trailing features are observed in some of the burst profiles, consistent with the characteristic signature of scattering. To estimate the scattering time, we adopt a one-component Gaussian function as the inherent shape of each burst. We allow the width of this Gaussian function to vary with frequency following the relationship $W_\nu = W_{\rm 1GHz} \cdot \nu^{\beta}$, where $W_{\rm 1GHz}$ represents the intrinsic pulse width at 1~GHz and $\beta$ is the exponent index that characterizes how the scattered pulse changes with frequency. In order to model the observed pulse profile, we use the convolution of a thin-screen pulse broadening function $H(\rm t) \cdot e^{-t/\tau}$ \citep{Cronyn70}. We concurrently fit for the scattering times ($\tau$) of different sub-bands utilizing the relationship $\rm \tau_\nu = \tau_{1GHz} \cdot \nu^{-4.0}$ \citep[e.g.,][]{nab13}. This fitting process is performed using the \texttt{emcee} Python package, which relies on the Markov Chain Monte Carlo method \citep{fhl13}. The resulting parameters, including the intrinsic pulse width $W_{\rm 1GHz}$, scattering times $\tau_{\rm 1GHz}$ and the pulse width-evolving index $\beta$, are presented in Table~\ref{tab:sourcelist}. 

Trailing features in the profiles might also arise from unresolved components, and it's plausible that the intrinsic evolution of the pulse profile with frequency could contribute to this effect. In Appendix \ref{app:w-e}, we explore the frequency-dependency of the pulse width and pulse energy.

\begin{figure*}
    \centering
    \includegraphics[width=0.35\textwidth]{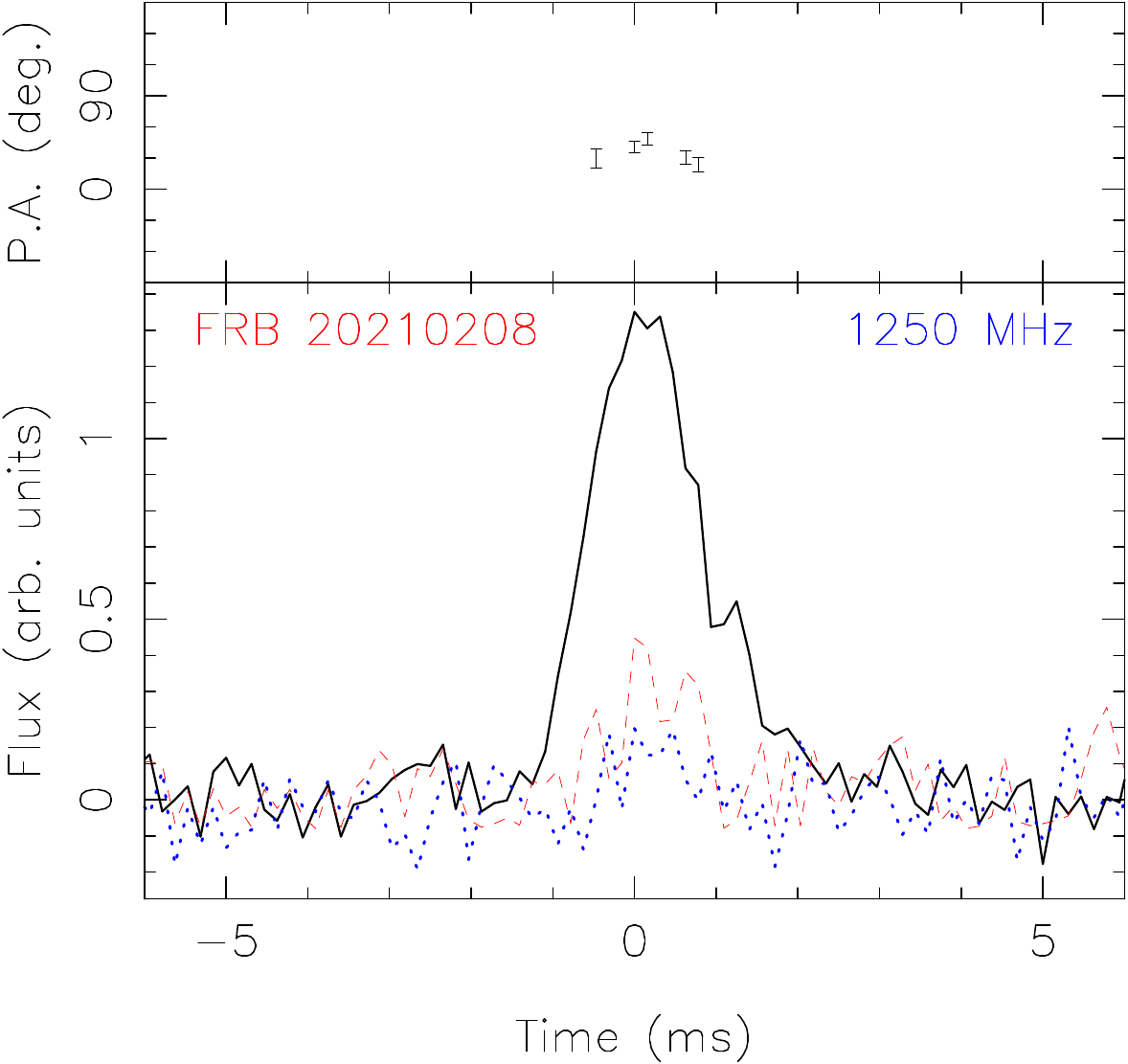} 
    \includegraphics[width=0.35\textwidth]{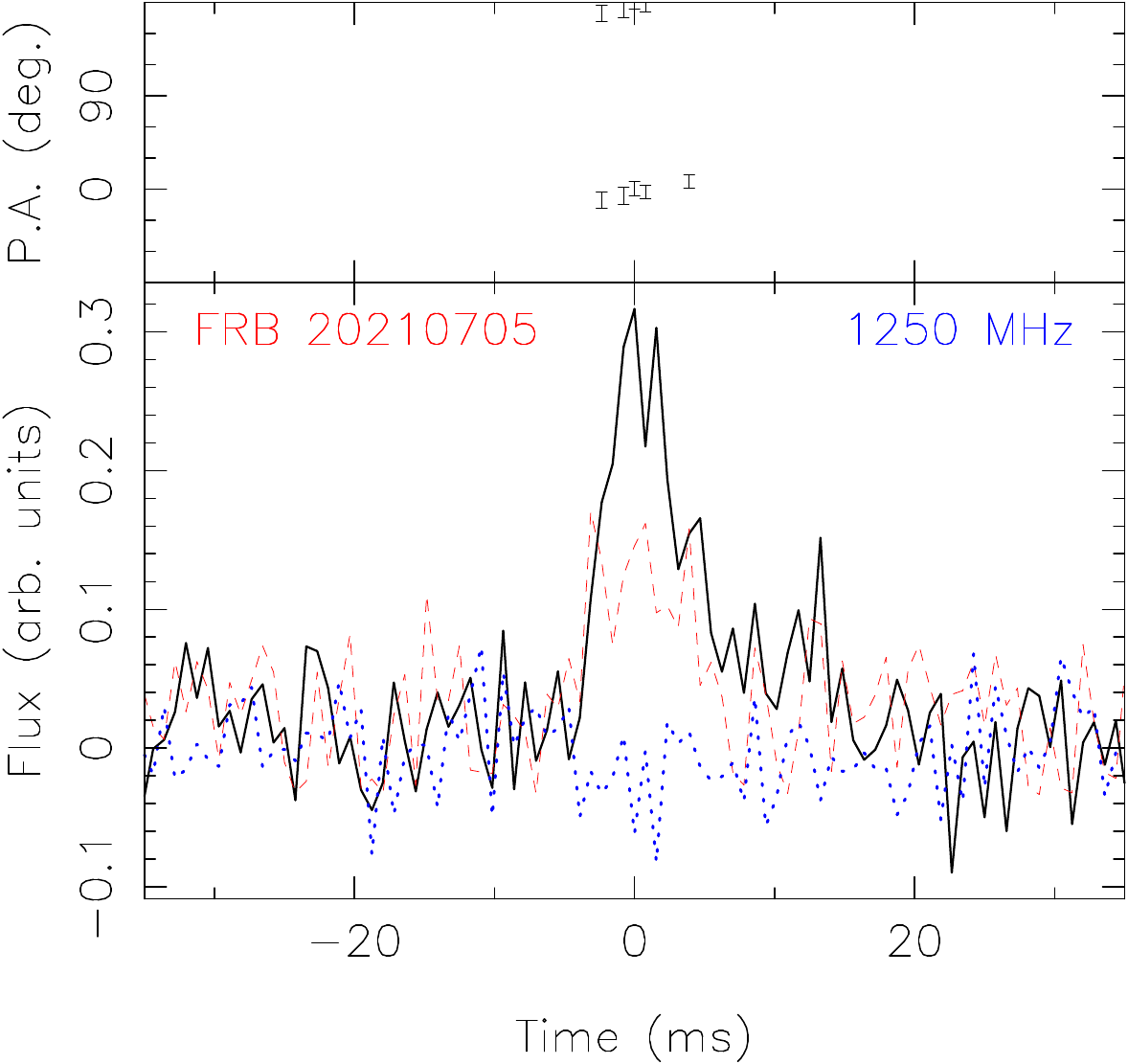} 
    \includegraphics[width=0.35\textwidth]{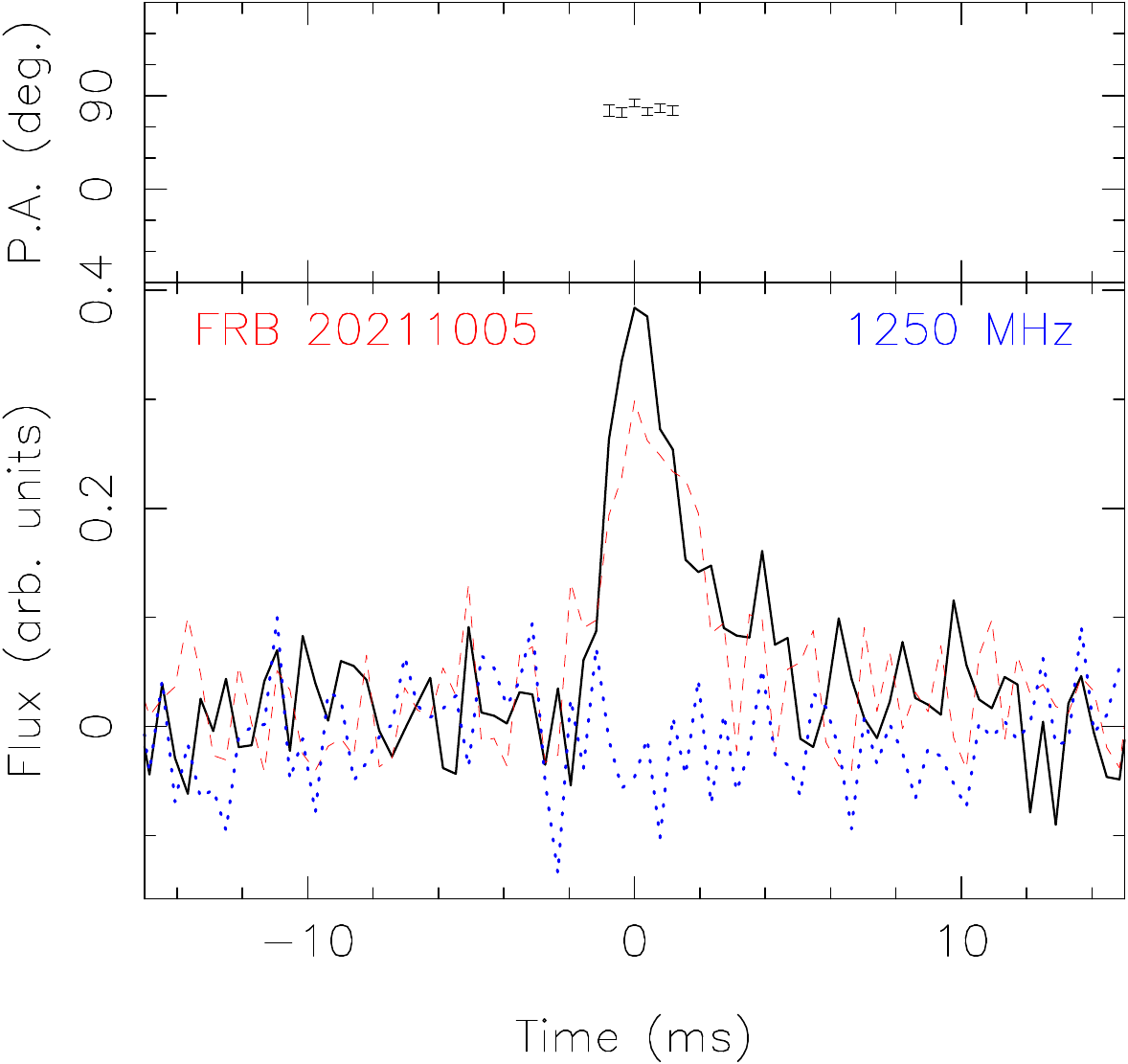} 
    \includegraphics[width=0.35\textwidth]{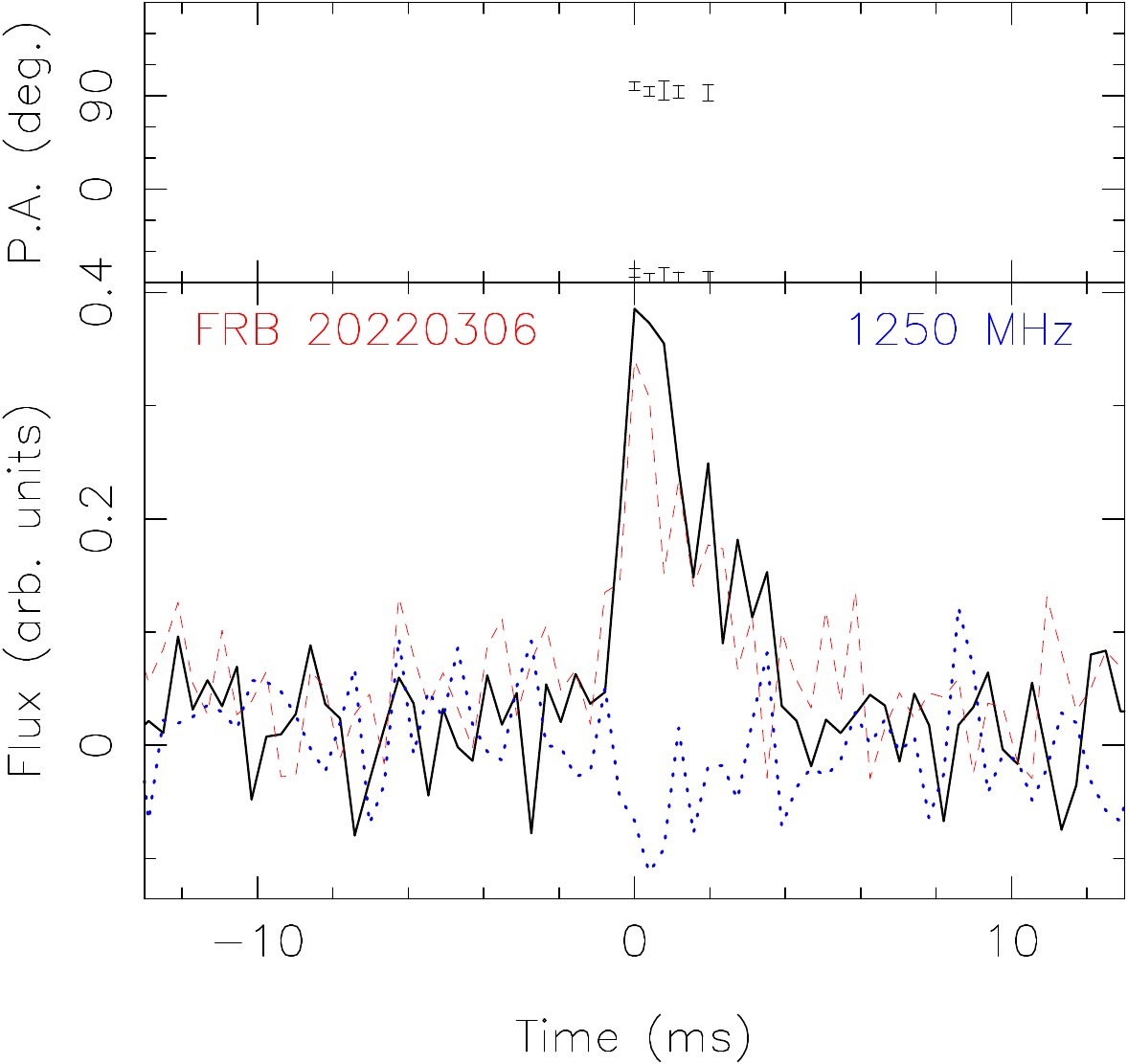}
    \caption{Polarization profiles of 4 FRBs discovered in the FAST GPPS survey. The top sub-panel is the polarization position angles $vs$ time. The bottom sub-panel shows the profiles of the Stokes $I$ (solid line), linear polarization $L$ (dashed line) and circular polarization (Stokes $V$, dotted line). The FRB name and central observation frequency are labeled.}
    \label{fig:polPlot}
\end{figure*}

\subsection{Peak flux density and fluence}

The flux density of each bin, denoted as $S_{\rm i}$ (measured in mJy), within the burst profile can be approximately calibrated using the standard deviation ($\sigma$) of the dedispersed data. This relationship is connected to the system noise as outlined in the work by \citet{zhou2023RAA}:
\begin{equation}
\sigma = \frac{T_{\rm sys} \cdot 10^{3}}{ G_0 \cdot \sqrt{ n_{\rm p} \cdot t_{\rm bin} \cdot BW}},
\end{equation}
here ${T_{\rm sys}}$ denotes the system noise temperature of 22~K. The effective gain of the telescope, $G_{\rm 0} = 16.1$ K/Jy, is also defined by \citet{JiangP2020RAA}. The term $n_{\rm p}=2$ represents the number of polarization channels summed, $t_{\rm bin}$ denotes the sampling time of the dedispersed data that is 196.608 $\mu$s, and $BW$ corresponds to the emission frequency bandwidth (in MHz). These FRBs have broadband radiation in general, and therefore we take the emission bandwidth as being the observation bandwidth of 437.5 MHz (i.e. 500~MHz subtracting the two band edges). Note that the energy at low frequencies of FRB\,20210705 have been scattered, and we have to take the data of the upper half for calculation. No emission is detected from FRB\,20210126 at high frequencies, possibly due to a very steep spectrum, and we have to take data of the lower half of the band. For these two FRBs, we take the bandwidth of 300 MHz in calculation. The results on flux density are listed in Table~\ref{tab:sourcelist}. 

The fluence $F_{\rm \nu}$ of a burst, measured in units of mJy\,ms at the central frequency $\nu=\rm1.25\,GHz$ (except for $\nu=\rm1.15\,GHz$ for FRB\,20210126 and $\nu=\rm 1.35\,GHz$ for FRB 20210705), is determined by integrating the previously defined flux density $S_{\rm i}$ over the sampling time $t_{\rm bin}$ (measured in ms). Mathematically, it can be expressed as $F_{\rm\nu} = \sum S_{\rm i} \cdot~t_{\rm bin}$. The uncertainties associated with the fluence and the peak flux density $S_{\rm peak}$ are calculated using the off-pulse noise as a reference. The results for fluence are listed in Table~\ref{tab:sourcelist}. 

We perform comparisons in Figure~\ref{fig:wf} between the $DM$, $S_{\rm peak}$, $F_{\rm \nu}$, $W_{\rm obs}$ and $DM_{\rm E}$ of one-off FRBs detected by FAST and those detected by other telescopes \citep{Keane2016Natur.530..453K,Caleb2017MNRAS.468.3746C,Petroff2017MNRAS,Bhandari+18MN,Patel2018ApJ...869..181P,Shannon2018Natur.562..386S,Bannister2019Sci...365..565B,Fedorova2019ARep,Macquart2019ApJ...872L..19M,Os2019MNRAS.488..868O,Connor2020MNRAS.499.4716C,Zhang2020ApJS..249...14Z,CHIME2021ApJS,Yu2022ATel15342....1Y}. Data from other telescopes are obtained from the FRB catalogue website\footref{fn:frbcat}. The FRBs discovered by FAST exhibit lower $S_{\rm peak}$ and $F_{\rm \nu}$ values. However, they display larger $DM$ and $DM_{\rm E}$ in comparison to FRBs discovered by other telescopes.

In Appendix \ref{app:w-e}, we compute the spectral index of a burst using the fitted fluences at various sub-bands. The spectral index is expressed as a function of $F_{\nu}=F_{\rm 1GHz} \cdot (\nu/{\rm1 GHz})^{\alpha}$, where $\alpha$ is the spectral index derived from the observed emission. The calculated spectral index values are listed in Table~\ref{tab:sourcelist}.

\subsection{Polarization}

Within the GPPS project, polarization data are typically collected during the follow-up FAST verification observations. When a previously known pulsar is located in a GPPS survey field, polarization data are also recorded in the survey observations. Fortunately, among the 5 FRBs being studied, two of them have their polarization data recorded during the GPPS survey observations, and the other two have their polarization data acquired during the GPPS follow-up tracking observations. The resulting polarization profiles are shown in Figure~\ref{fig:polPlot}.

Based on detected linear polarization, the RMs of the four FRBs are also determined. These RMs encompass contributions from the Milky Way ($RM_{\rm MW}$), the intergalactic medium ($RM_{\rm IGM}$) and the host galaxy ($RM_{\rm host}$). The contribution of the Galactic RM can be estimated by calculating the median RM of the nearest 20 background extragalactic radio sources \citep{Xu2014RAA}. After the RM foreground from the Milky Way is discounted, the left is the extragalactic Faraday rotation measures of the FRBs, $RM_{\rm E}$, which include the contributions from the host galaxy and the intergalactic medium. In general, the RMs from the intergalactic medium are small \citep{xuhan14,xuhan22,car22}, and such a large $RM_{\rm E}$ probably should be attributed to the host galaxy. No doubt that $DM_{\rm E}$ probably mostly comes from the intergalactic medium, then the  $DM_{\rm E}$ can be taken as the upper limit of the contribution from the host galaxy, then the {\it lower limit} of magnetic field along the line of sight in the host galaxy could be $\langle B \rangle_{\rm E} = 1.232 \times RM_{\rm E} /DM_{\rm E}$. Particularly, FRB\,20211005 exhibits the highest absolute RM of $1739\pm4\rm\,rad\,m^{-2}$ and an $RM_{\rm E}$ of $1661\pm30\rm\,rad\,m^{-2}$, which may indicate a complex environment in the host galaxy.

\section{Discussion}
\label{4-conc}

From the FAST GPPS survey data, we have successfully discovered 5 new FRBs, adding to the 5 previously FAST-discovered FRBs reported by the CRAFTS project \citep{ZhuWW2020ApJ, NiuxCH2021ApJ, NiuCH2022Natur}. This brings the total count of FRBs discovered by FAST to 10. Considering the accumulated observation time of 17747 beam*hours and taking into account the FAST 19-beam L-band receiver with a beam size of $\rm 3^\prime$, one can compute the FRB detection rate within the GPPS survey. At the 95\% confidence level, the detction rate is approximately $\rm~5.3^{+5.2}_{-3.0}\times10^{-3}~hr^{-1}$, as per the method presented by \citet{Gehrels1986ApJ}. Comparatively, this rate is slightly higher than the FRB detection rate of $\rm3.0^{+3.3}_{-2.0}\times10^{-3}~hr^{-1}$ reported by the CRAFTS project \citep{ZhuWW2020ApJ, NiuxCH2021ApJ, NiuCH2022Natur}, and also surpasses the previous estimate of $2.5^{+7.9}_{-2.3}\times 10^{-3},\rm hr^{-1}$ by \citet{Luo2020MNRAS}. The results suggest that FAST may detect more FRBs than previously anticipated, though the rates are consistent with each other within uncertainties.

As illustrated in Figure~\ref{fig:wf}, the FRBs detected by the FAST exhibit remarkably low peak flux densities and fluences. However, they display substantial DMs, along with pronounced $DM_{\rm E}$ values on a broader scale. The derived parameters from our comprehensive analysis of these five FRBs, which were identified within regions of low Galactic latitude, offer valuable constraints for our Milky Way. 

In addition, our FRB searching in an extended range of DMs from 3600$\rm~pc\,cm^{-3}$ to 10000$\rm~pc\,cm^{-3}$ (also with a step of 1.0$\rm~pc\,cm^{-3}$) also indicates a lack of positive outcomes. This absence of results can be understood since larger DMs might suggest higher redshifts and correspondingly increased cosmological distances, as well as heightened levels of scattering within the frequency band observed by FAST.

\section*{Acknowledgements}
The anonymous referee is acknowledged for his very careful readings and very thoughtful suggestions.
The GPPS survey, as one of five key projects, is being carried out by using FAST, a Chinese national mega-science facility built and operated by the National Astronomical Observatories, Chinese Academy of Sciences.
J.~L. Han is supported by the National Natural Science Foundation of
China (NSFC, Nos. 11988101 and 11833009) and the Key Research
Program of the Chinese Academy of Sciences (Grant No. QYZDJ-SSW-SLH021);
D.~J. Zhou is supported by the Cultivation Project for the FAST
scientific Payoff and Research Achievement of CAMS-CAS.
%
C. Wang, P.~F. Wang
are supported by NSFC
No.~12133004 and 
also partially
supported by the National SKA program of China No.~2020SKA0120200.
In addition, 
C. Wang is also partially supported by NSFC No.~U1731120;
P.~F. Wang is also partially supported by the NSFC No.~11873058;
Jun Xu is partially supported by the National SKA Program of China (Grant No. 2022SKA0120103) and  NSFC No.~U2031115.
%
H.~G. Wang is supported by NSFC No.~12133004, and Guangzhou Science and Technology Project No.~202102010466.

\section*{Data Availability}

Original FAST observational data will be open sources according to the FAST data 1-year protection policy. The processed data shown in this paper can be obtained from authors by kind requests.

\bibliographystyle{mnras}
\bibliography{bibfile} 

\appendix
\section{Evolution of Pulse width and spectra with frequency} 
\label{app:w-e}

We investigate the frequency evolution of pulse width and pulse energy by fitting the pulse width and fluence against frequency by the MCMC. 

\begin{figure*}
    \centering
    \hspace{6mm}
    \includegraphics[width=0.30\textwidth]{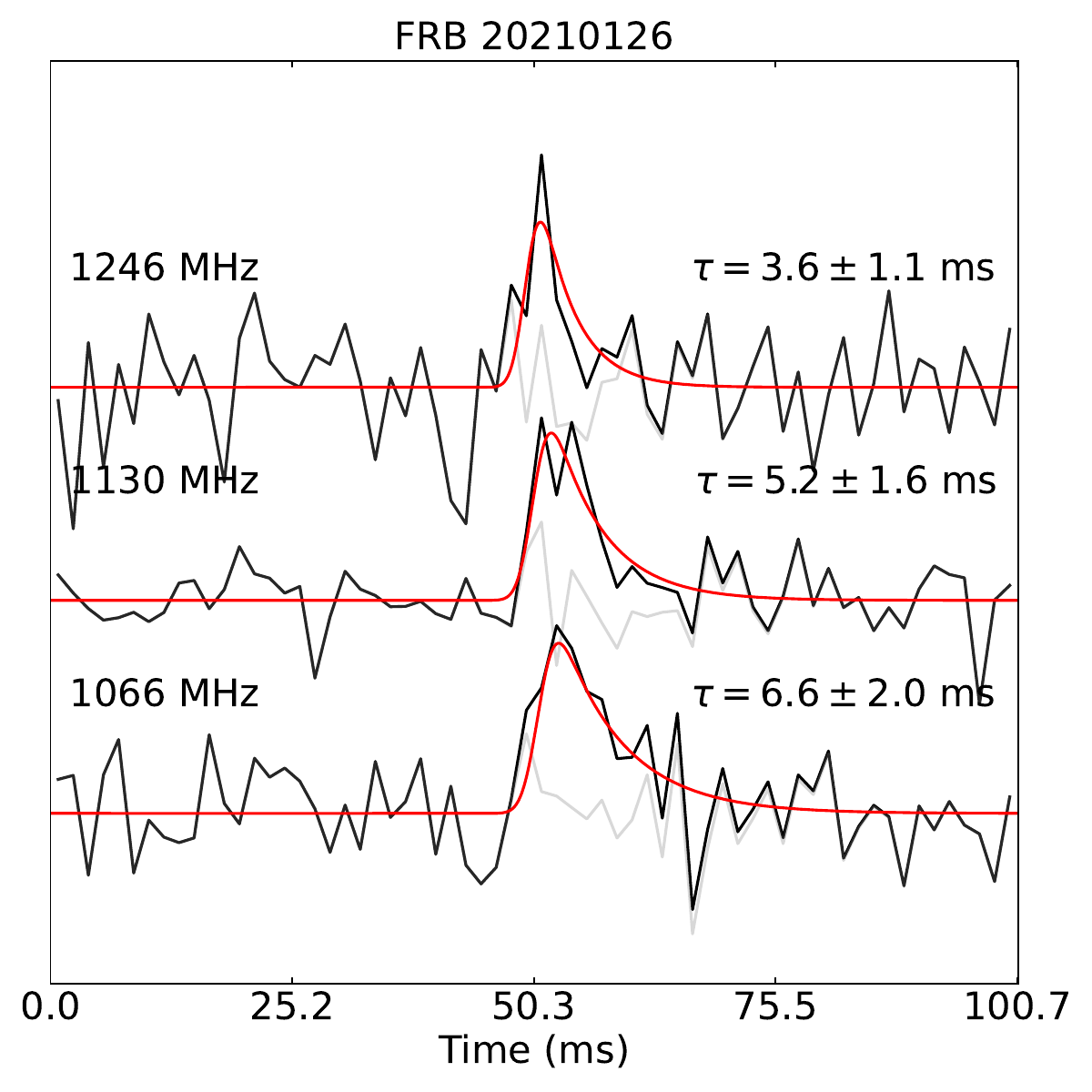}
    \hspace{1mm}
    \includegraphics[width=0.30\textwidth]{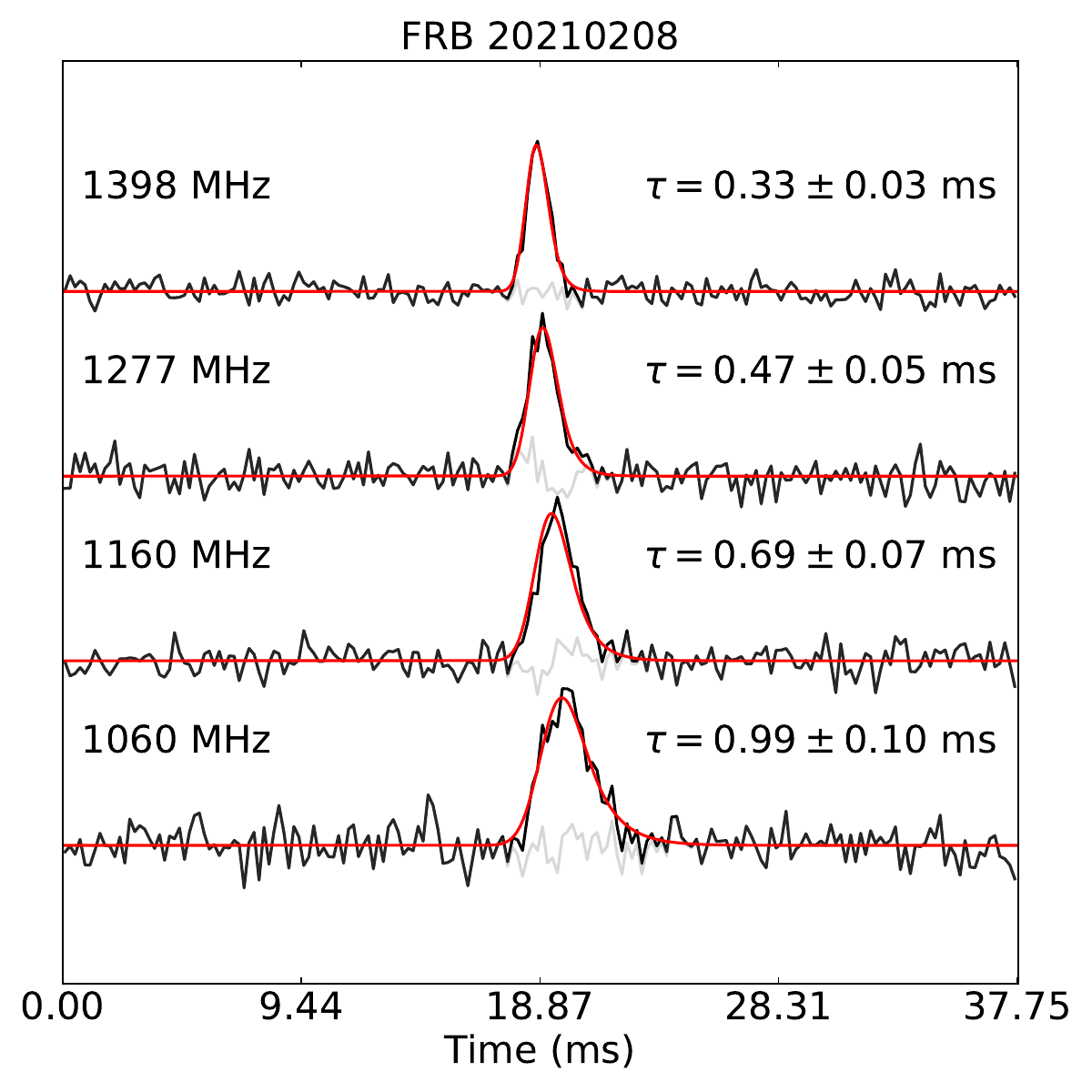}
    \hspace{1mm}
    \includegraphics[width=0.30\textwidth]{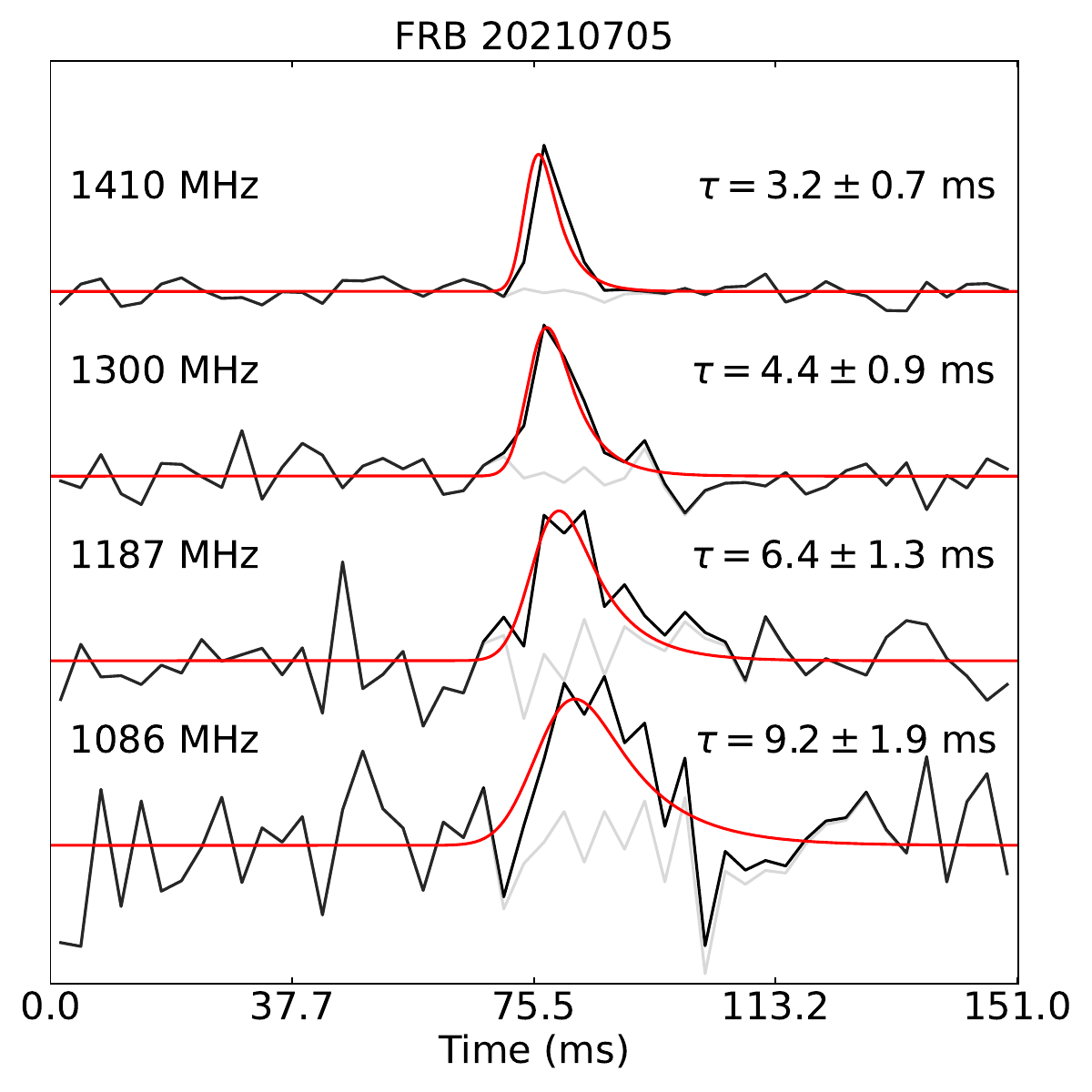} \\
    \includegraphics[width=0.31\textwidth]{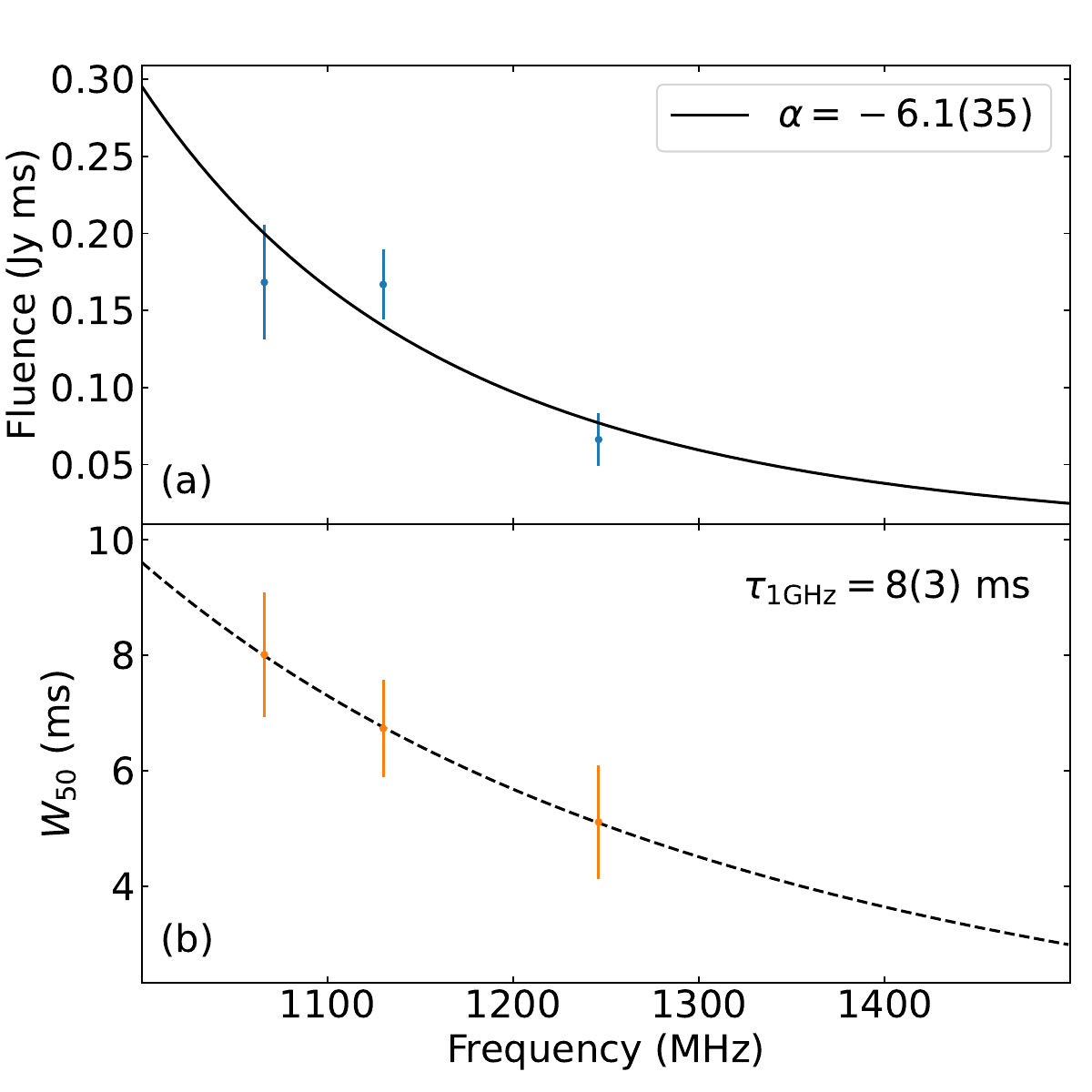}
    \includegraphics[width=0.31\textwidth]{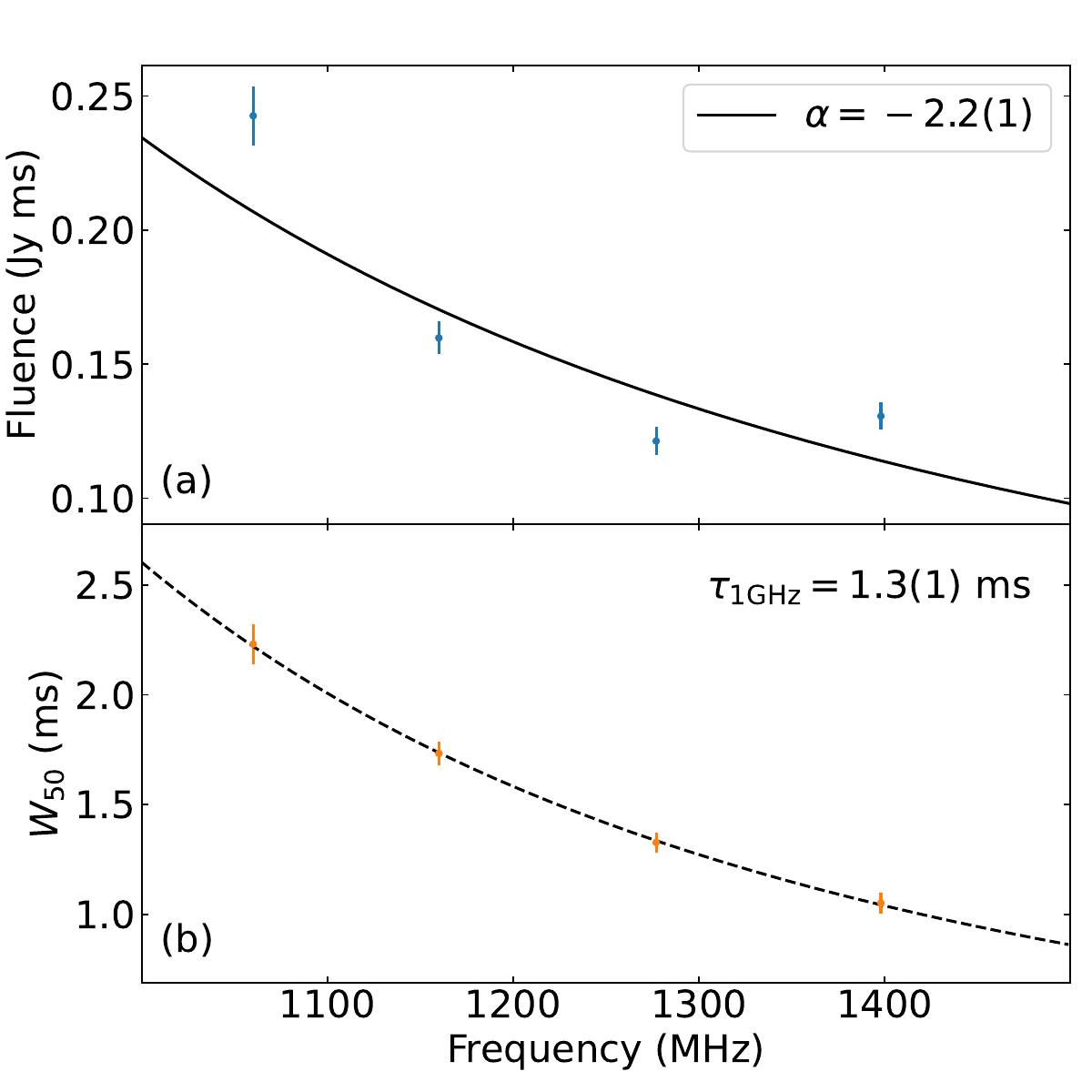}
    \includegraphics[width=0.31\textwidth]{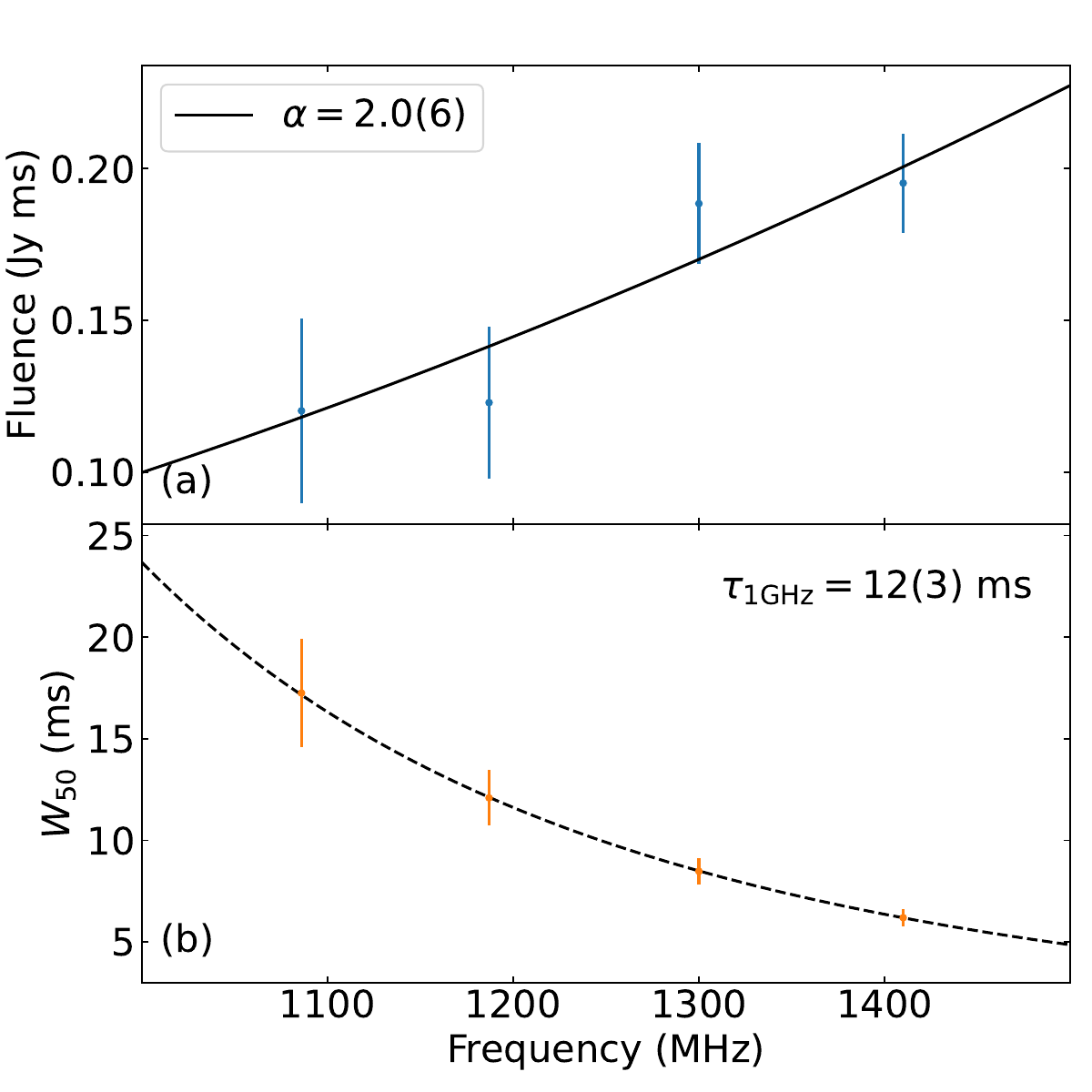} \\
    \hspace{4mm}
    \includegraphics[width=0.30\textwidth]{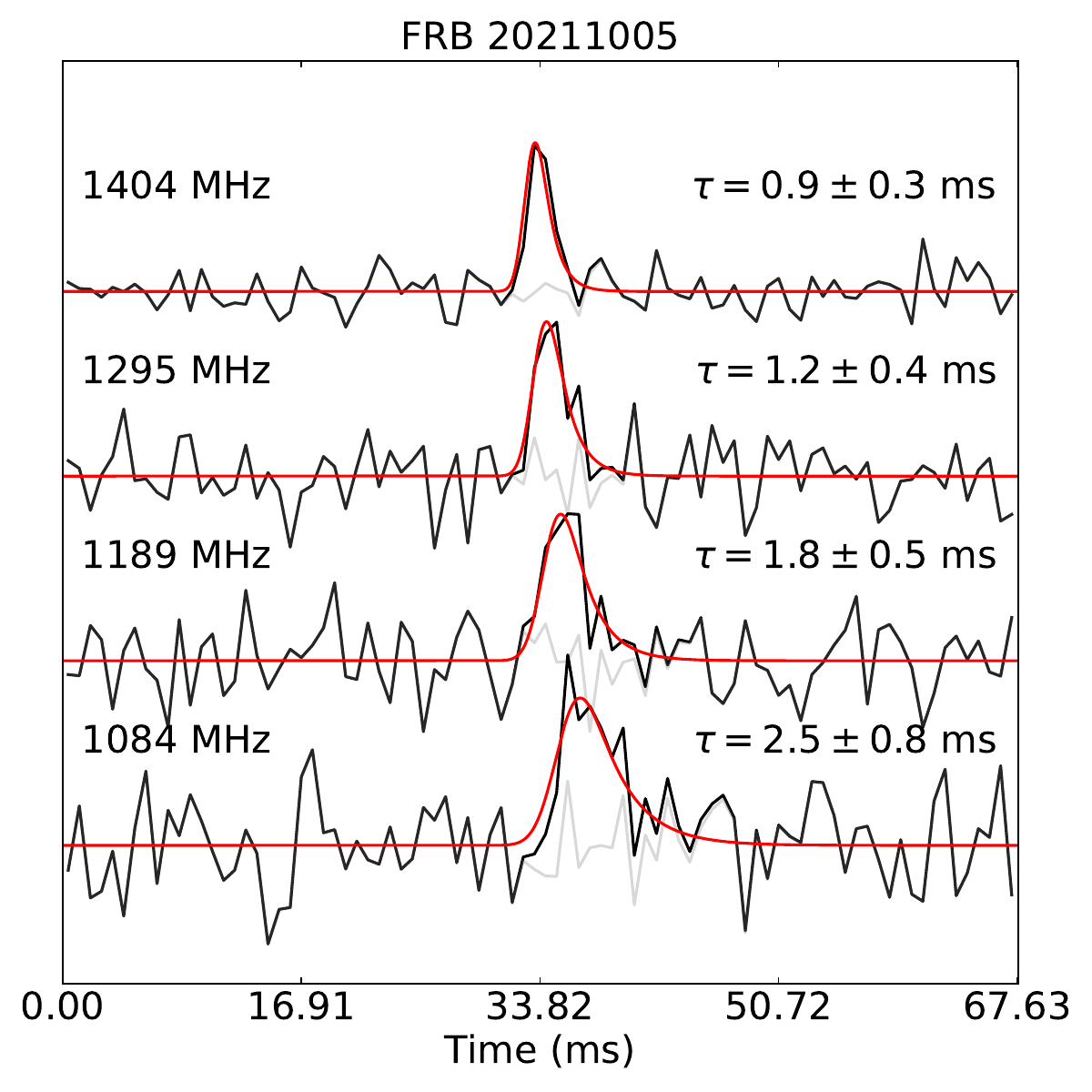}
    \hspace{1mm}
    \includegraphics[width=0.30\textwidth]{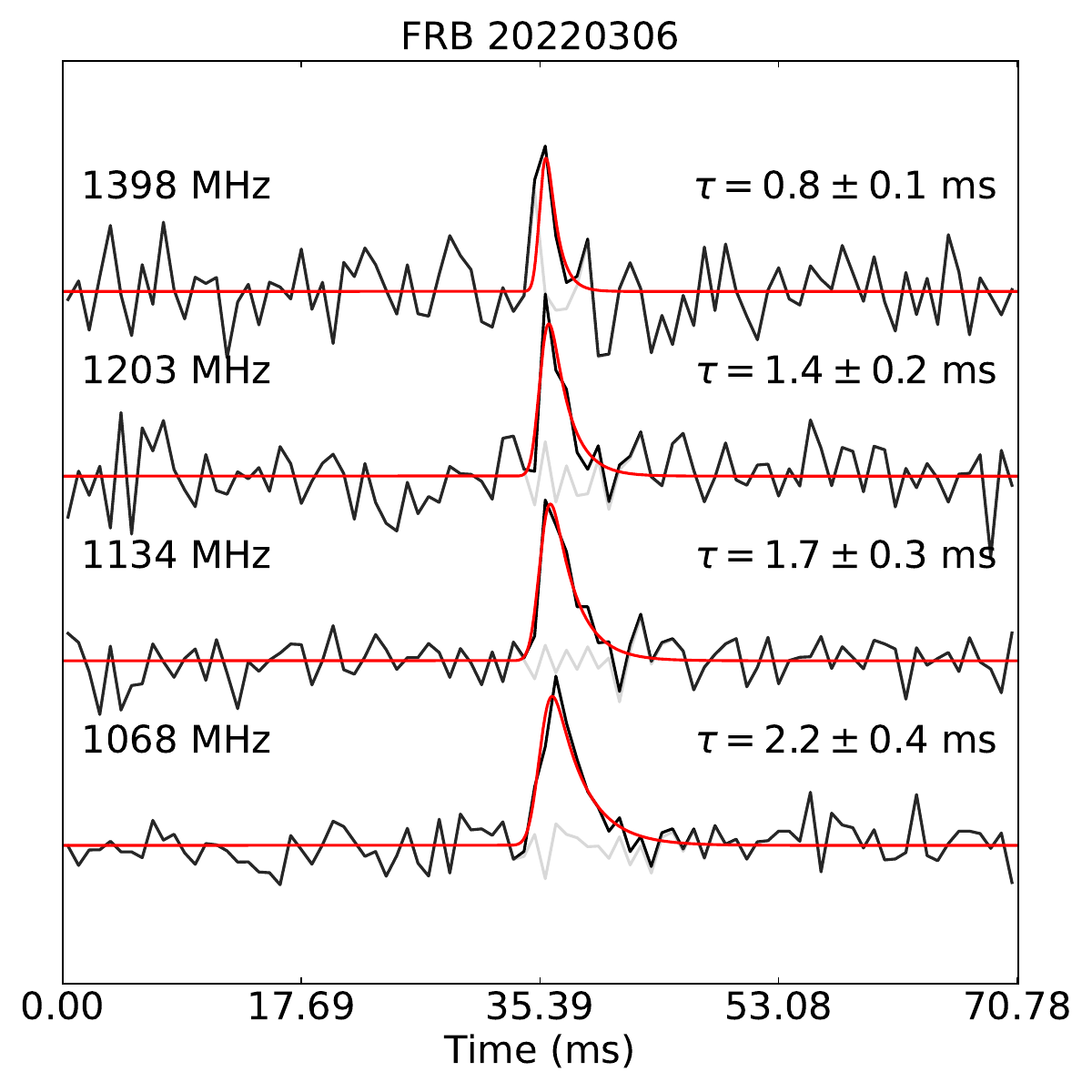} \\
   \includegraphics[width=0.31\textwidth]{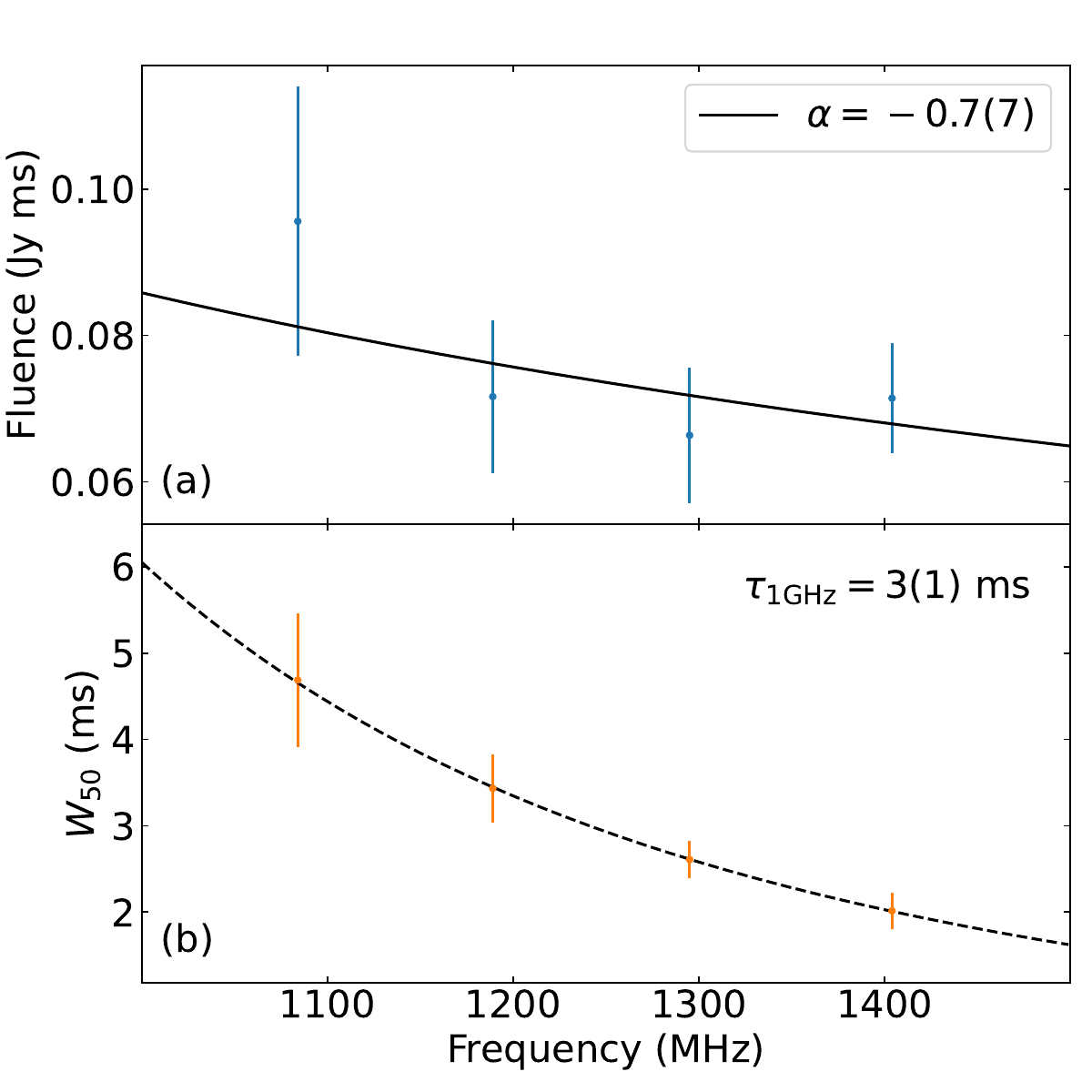}
   \includegraphics[width=0.31\textwidth]{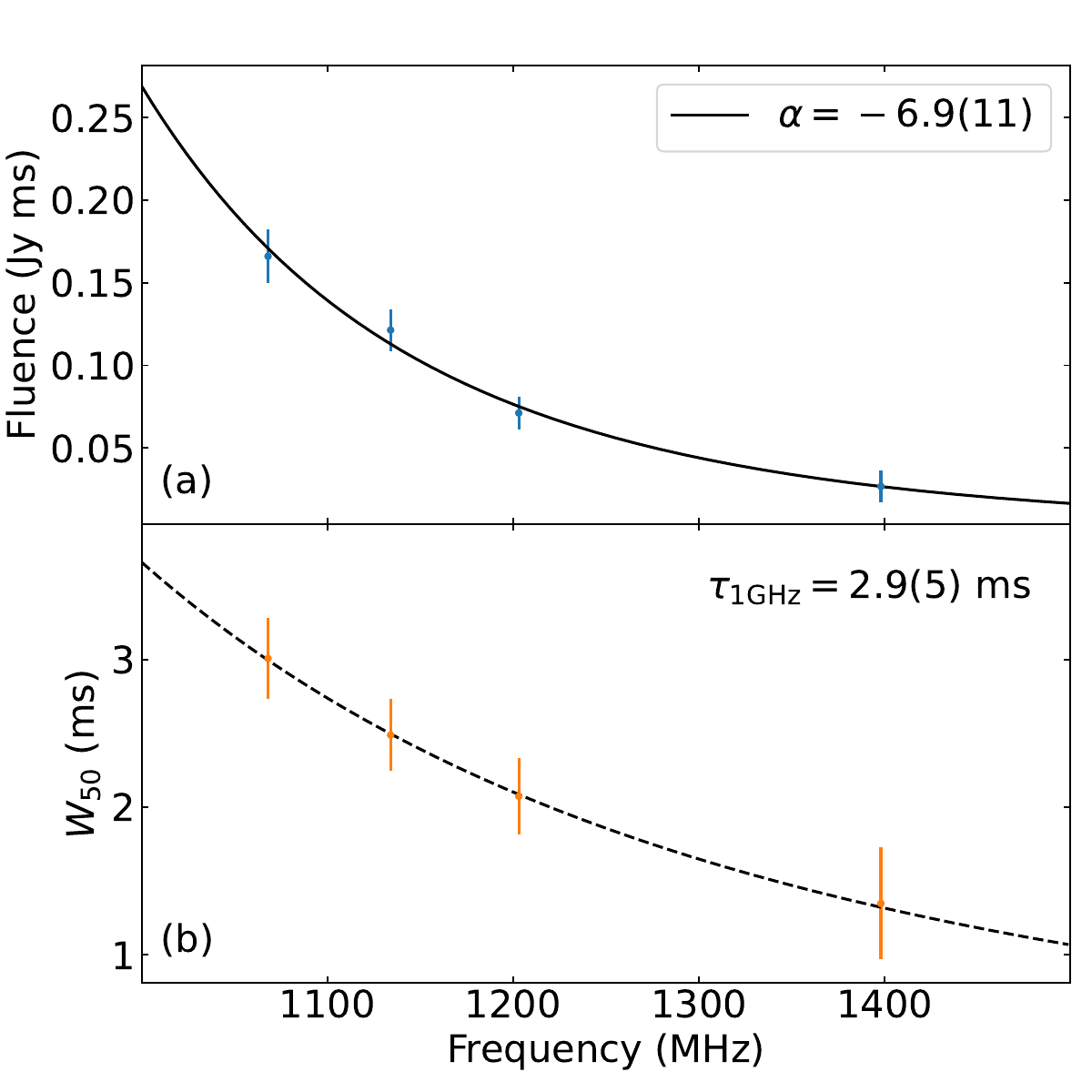}
   \caption{For each FRB, the upper panel shows the observed subband profiles and their fitted results by the MCMC, with the fitted subband scattering timescale marked on the right. In the subpanel (a) of the lower panel the observed subband fluences (with error bars) evolve with frequency, and are fitted by equation $F_{\nu}=F_{\rm 1GHz} \cdot (\nu/{\rm1 GHz})^{\alpha}$. In the sub-panel (b), the pulse widths of subbands (with error bars) are fitted by the MCMC, with the scattering time scale $\tau(\nu)=\tau_{\rm 1GHz} \cdot \nu^{-4}$.
    }
    \label{fig:20210126}
\end{figure*}

First of all, we get the subband profiles and fit the profiles to get the results as shown in Figure \ref{fig:20210126} for each FRB. The pulse width (full-width at half maximum, FWHM) of each subband profile is estimated from the fitted profiles. 
Second, we fit the pulse width with the scattering time scale $\rm \tau_\nu = \tau_{1GHz}\cdot\nu^{-4.0}$ in the MCMC for each FRB.  We also fit the pulse energy for the spectral index.

\label{lastpage}
\end{document}